\documentclass[aps,pre,twocolumn,superscriptaddress,longbibliography,nobibnotes,nodoi,nourl,noeprint]{revtex4-2}
\usepackage{amsmath,amssymb,physics,bm,siunitx}
\usepackage{xcolor,graphicx,soul}
\usepackage[colorlinks=true,citecolor=blue,urlcolor=blue,linkcolor=black]{hyperref}

\def\rms{\rm\scriptscriptstyle}
\newcommand{\expo}{p}
\newcommand{\Cpp}{C\nolinebreak\hspace{-.05em}\raisebox{.4ex}{\tiny\bf +}\nolinebreak\hspace{-.10em}\raisebox{.4ex}{\tiny\bf +}}
\DeclareMathOperator*{\argmin}{argmin}

\begin{document}
	
\title{Brownian dynamics simulations of hard rods in\\ external fields and with contact interactions}

\author{Alexander P.\ Antonov}
\email{alantonov@uos.de}
\affiliation{Universit\"{a}t Osnabr\"{u}ck, Fachbereich Physik, Barbarastra{\ss}e 7, D-49076 Osnabr\"uck, Germany}

\author{S\"{o}ren~Schweers}
\email{sschweers@uos.de}
\affiliation{Universit\"{a}t Osnabr\"{u}ck, Fachbereich Physik, Barbarastra{\ss}e 7, D-49076 Osnabr\"uck, Germany}

\author{Artem Ryabov}
\email{rjabov.a@gmail.com}
\affiliation{Charles University, Faculty of Mathematics and Physics, Department of Macromolecular Physics, V Hole\v{s}ovi\v{c}k\'{a}ch 2, 
CZ-18000 Praha 8, Czech Republic}

\author{Philipp Maass} 
\email{maass@uos.de}
\affiliation{Universit\"{a}t Osnabr\"{u}ck, Fachbereich Physik, Barbarastra{\ss}e 7, D-49076 Osnabr\"uck, Germany}

\date{November 15, 2022} 

\begin{abstract}
We propose a simulation method for Brownian dynamics of hard rods in one dimension for arbitrary continuous 
external force fields. It is an event-driven procedure based on the fragmentation and mergers of clusters formed 
by particles in contact. It allows one to treat particle interactions in addition to the hard-sphere exclusion as long as the corresponding interaction forces are continuous functions of the particle coordinates. We furthermore develop a treatment of 
sticky hard spheres as described by Baxter's contact interaction potential.
\end{abstract}

\maketitle

\section{Introduction}

Motion of micron-sized particles in soft matter and biological systems can in general be well
described by an overdamped Brownian motion.
In these dynamics, the excluded volume part of the particle-particle interactions is often taken into account by considering the particles as
hard spheres. The hard-sphere interaction represents a core part of any interaction if particles cannot penetrate each other and has a decisive impact on collective phenomena. Even if a softcore potential is more accurately accounting for the repulsive interactions
when particles approach each other, the hard-sphere model is often used as an effective description. However, in Brownian dynamics simulations the hard-sphere interaction requires special care due to its singular nature.

The problem of simulating hard spheres has been around as long as modern computational physics \cite{Rosenbluth/Rosenbluth:1954, Alder/Wainwright:1957}. In these early studies, equilibrium properties were simulated based on Monte Carlo 
methods. The first dynamic algorithm was similar to standard Monte Carlo simulations and based on rejecting movements when 
particle positions violate the hardcore constraint \cite{Cichocki/Hinsen:1990}. A modification to this approach was made by placing 
particles in contact instead of rejecting the movements \cite{Schaertl/Sillescu:1994}. However, these first proposed methods are not suitable for 
high particle densities. Recent developments have shown that Monte-Carlo schemes can give
quantitative agreement with Brownian dynamics simulations even at high densities \cite{Jabbari-Farouji/Trizac:2012, Corbett/etal:2018, Daza/etal:2020}.

An alternative means of simulating hardcore particle interaction are event-driven algorithms, where particle overlappings are
 avoided, similarly as it is often done in molecular dynamics simulations \cite{Hurtado/Garrido:2020}.
Most of these approaches are based on the idea that when an attempted displacement of the particles in the numerical procedure leads to an overlap  of two particles, the  corresponding encounter of the two particles is treated like a binary elastic collision.
For treating a case where more than two particles overlap, a sequence of binary collision can be carried out.
For zero external forces, a corresponding procedure was developed in Refs.~\cite{Strating:1999, Scala:2012}, and for non-zero external forces a refined method was proposed in Ref.~\cite{Behringer/Eichhorn:2012}.
One may also consider binary collisions which are not perfectly elastic \cite{Noije/etal:1998, Luding:1998, Gonzalez:2009}. 

In real systems, it is often found that more than two particles come into contact \cite{Helland/etal:2007, McMillan/etal:2013, Cahyadi:2017}. Then, particles can be grouped into clusters. The presence of clusters can be justified by kinetic theory, thereby establishing the limits of applicability of the binary collision theory (inelastic or elastic) \cite{Lois/etal:2006}. Aggregation of such clusters in Brownian dynamics simulation was described in \cite{Watanabe/Tanaka:2013}.

Here we present a method for simulating overdamped Brownian dynamics of hard rods in one dimension based on particle clusters. It relies on an exact method to solve the Langevin equation for this many-body system in arbitrary continuous external force fields. Particle configurations are evolved such that particle clusters are moved coherently. The clusters are formed by particles that are in contact. They can fragment into smaller or merge into larger clusters, and the fragmentation and merging processes are determined by conditions on the mean external forces acting on the clusters. 
	
In several applications, in addition to excluded volume interactions, attractive interaction forces need to be taken into account.
To this end, a simple generalization of the hard-sphere model was introduced by Baxter \cite{Baxter:1968}. In this model of sticky hard spheres, adhesive interactions are present when particles are in contact. Equilibrium properties of this model
can be studied analytically in one dimension \cite{Percus:1982}. Beyond one dimension, structural properties 
were studied extensively by Monte Carlo simulations in the past \cite{Seaton/Glandt:1987_2,Miller/Frenkel:2004_2,Miller/Frenkel:2004_1, Foffi/etal:2005, Zaccarelli:2007, Wang/Swan:2019, Whitaker/etal:2019, Stopper/etal:2019, Bou-Rabee/etal:2020, Holmes-Cerfon:2020}.  Approaches based on rates for contact breaking and
formation between spheres were developed to describe nonequilibrium kinetics of such systems \cite{Babu/etal:2006}.

For including the contact interaction into Brownian dynamics simulations, we present a way of representing the $\delta$-singularity in the sticky hard-sphere interaction potential. This allows one to tackle the contact interaction in our cluster-based algorithm. One can use this method also in other simulation procedures for the Brownian dynamics of hard spheres.

\section{Langevin equations for Brownian motion of hard rods}
\label{sec:algorithm}
The one-dimensional overdamped Brownian motion of $N$ hard rods of length $\sigma$ with center positions at $x_i$, $i=1,\ldots, N$, 
in an external force field $f(x)$ is described by the Langevin equations
\begin{equation}
\frac{\dd x_i}{\dd t}=\mu f(x_i)+\sqrt{2D}\,\xi_i(t),
\label{eq:langevin}
\end{equation}
where $\mu$ is the particle mobility, $D$ is the diffusion coefficient, and $\xi_i(t)$ are Gaussian white noise processes with 
zero mean and correlation functions $\langle \xi_i(t) \xi_j(t') \rangle = \delta_{ij}\delta(t - t')$. 

If the particles would exert continuously varying interactions forces on each other, one could add these forces to the
right hand side of  Eqs.~\eqref{eq:langevin} and simulate the particles' motion by standard methods \cite{Kloeden/Platen:1992, Saito/Mitsui:1993}, 
e.g.\ by the Euler algorithm in the simplest approach. For hard rods, however, the interaction cannot be described by a continuously varying force. It enters the dynamics as a condition imposed on probability currents in the many-particle Fokker-Planck equation (see supplementary material of Ref.~\onlinecite{Lips/etal:2018}). The accurate and efficient treatment of this interaction in the Langevin equations \eqref{eq:langevin} needs special care. 

In the following we present a method to solve Eqs.~\eqref{eq:langevin}, which we refer to as 
Brownian cluster dynamics (BCD) simulations. 
We will also consider Baxter's sticky hard-sphere interaction \cite{Baxter:1968}, 
which corresponds to an additional attraction
between two hard rods when they get in contact.
	
\section{BCD simulations for hard rods in external force field}
\label{sec:algorithm}

Our new algorithm 
gives an approximate solution of the Langevin equations \eqref{eq:langevin} by evolving the system in fixed time steps $\Delta t$.
It is based on conditions for cluster movements. Such movements are highly relevant, for example, to understand unexpected particle currents
in driven Brownian motion through highly populated periodic potentials \cite{Antonov/etal:2022a}.
A cluster of size $n$
is a local arrangement of $n$ rods that are mutually in contact, but not in contact with other rods. 
We call such an arrangement an $n$-cluster. Single rods in this description are 1-clusters.

In dense systems, there will be clusters of all sizes that can fragment into smaller ones and/or collide with neighboring clusters during their motion. The challenge is to identify these fragmentation and collision events accurately. 

We do this by first identifying at a starting time $t$ how the clusters move.
This ``cluster analysis'' takes into account the fragmentation events. It amounts to assigning a velocity to each particle at time $t$, where particles moving together as a cluster have the same velocity. An attempt is then made to propagate the particle positions with these velocities in a small time step $\Delta t$. If there is no collision of clusters in the interval $[t,t+\Delta t[$, the particles are displaced correspondingly, and the cluster analysis is carried out for the new particle configuration at time $t+\Delta t$. 

In the case when there is a collision of clusters during the interval $[t,t+\Delta t[$, the first collision event 
at time $t_{\rm c}$ is determined. For the clusters involved in that collision, the cluster analysis is carried out, yielding updated velocities for the particles being part of the colliding clusters. Then again it is attempted to propagate the particle positions with the fixed velocities for all particles
until time $t+\Delta t$.
If there is a further collision, this is taken into account in the same manner as the first collision. 

The process is repeated until the time $t+\Delta t$ is reached. Then all particle velocities are updated by a cluster analysis. 
	
\subsection{Cluster analysis}
We consider $n$ hard rods of length $\sigma$ forming an $n$-cluster at time $t$. 
The particle positions in the $n$-cluster are ordered from left to right,
\begin{equation}
x_2=x_1+\sigma,\, \ldots,\, x_n=x_1+(n-1)\sigma\,,
\end{equation}
where $x_1$ is the position of the leftmost particle. In the cluster, the particles
can exert forces on each other that must obey the following conditions:
\begin{list}{--}{\setlength{\leftmargin}{1.2em}\setlength{\rightmargin}{0em}
\setlength{\itemsep}{0ex}\setlength{\topsep}{0.5ex}}
\item The interaction force $F^{\rm int}_{i,i+1}$ of particle $i$ on particle $(i\!+\!1)$ must be non-negative, $F^{\rm int}_{i,i+1}\ge0$.
\item The force $F^{\rm int}_{i+1,i}$ exerted by particle $(i\!+\!1)$ on particle $i$ is 
$F^{\rm int}_{i+1,i}=-F^{\rm int}_{i,i+1}$.
\end{list}
We furthermore define by 
\begin{equation}
F^{\rm ext}_i = f(x_i) + \, \frac{\sqrt{2D}}{\mu}\,\xi_i
\label{eq:fexttot}
\end{equation}
the total external force,
including the stochastic force mediated by the surrounding fluid.

The particles in the $n$-cluster can move in different ways. 
For example, in case of a 3-cluster, all particles can move as single particles (1-clusters).
Or only the left particle moves as a single particle (1-cluster), while the
middle and right particle move together as a 2-cluster. Or the left and middle particle move as a 2-cluster, and the right one as a 1-cluster. Or all 3 particle keep in touch and the 3-cluster moves as a whole. We can distinguish between these possibilities by considering the 4~compositions 
$\{1,1,1\}$, $\{1,2\}$, $\{2,1\}$, and $\{3\}$ of the 3-cluster into the respective subclusters. 

\begin{figure}[t!]	
\centering
\includegraphics[scale=1]{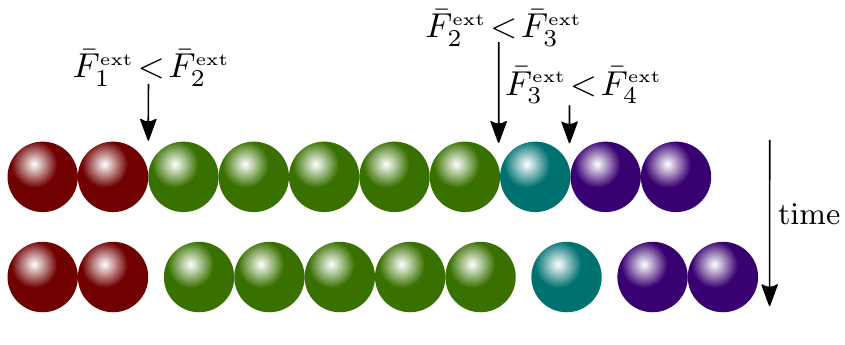}
\caption{Illustration of a possible fragmentation of a 10-cluster into a 2-, 5-, 1-, and 2-cluster.
The conditions for the external forces giving rise to such fragmentation are given 
in Eqs.~\eqref{eq:cond1} and \eqref{eq:cond2}.}
\label{fig:illustration_cluster_evolution} 
\end{figure}

For an $n$-cluster, there exist $2^{n-1}$ possible 
compositions (fragmentations) $\{m_1,\ldots,m_s\}$ into subclusters of sizes 
$m_1,\ldots,m_s$, with $1\le m_j\le n$, $\sum_{j=1}^s m_j=n$.
A possible fragmentation of a 10-cluster into 4 subclusters is illustrated in Fig.~\ref{fig:illustration_cluster_evolution} together with the conditions for this fragmentation to occur. The conditions for a general composition (fragmentation) $\{m_1,\ldots,m_s\}$ to occur, are explained next.

Let us first consider the condition for the $j$th subcluster of size $m_j$ to become separated from the $(j\!+\!1)$th subcluster of size $m_{j+1}$. All particles in a moving subcluster must have the same velocity, and this velocity must be the same as that of the center (of mass) of the subcluster. The velocity of the $j$th subcluster is 
equal to $\mu\bar F^{\rm ext}_j$, where 
\begin{equation}
\bar F^{\rm ext}_j=\frac{1}{m_j}\sum_{k=1}^{m_j} F^{\rm ext}_{j,k}
\label{eq:mean-Fext}
\end{equation}
is the mean external force exerted on the particles in this subcluster. 
Here,  $F^{\rm ext}_{j,k}$ is the external force on particle $k$ in the subcluster $j$, i.e.\ it is equal to
the force $F^{\rm ext}_l$ on particle $l$ in the $n$-cluster, where $l=k+\sum_{i=1}^{j-1} m_i$.  

%%%%%%%%%%%%%%%%%%%%%%%%%%%%%%%%%%%%%%%%%%%%%%%%%%
%%%%%%%%%%%%%%%%%%%%%%%%%%%%%%%%%%%%%%%%%%%%%%%%%%

For the $j$th  and $(j\!+\!1)$th  subclusters to separate, the velocity of the $(j\!+\!1)$th subcluster must be larger than that of the $j$th subcluster, i.e.\ the condition
$\bar F^{\rm ext}_j<\bar F^{\rm ext}_{j+1}$ must be fulfilled. The same holds true for the other subcluster separations, i.e.\ we obtain
\begin{equation}
\bar F^{\rm ext}_j<\bar F^{\rm ext}_{j+1}\,,\hspace{1em}j=1,\ldots,s-1\,.
\label{eq:cond1}
\end{equation}
as a first set of conditions for the composition $\{m_1,\ldots,m_s\}$ to occur.

In addition, the particles in the $j$th subclusters need to keep in contact. This means that the condition~\eqref{eq:cond1} 
for subcluster separation must be violated for all possible fragmentations of this subcluster. We thus obtain
\begin{equation}
\frac{1}{i}\sum_{k=1}^iF^{\rm ext}_{j,k}\ge 
\frac{1}{m_j-i}\sum_{k=i+1}^{m_j} F^{\rm ext}_{j,k}\,, \hspace{1em}i=1,\ldots,m_j-1\,.
\label{eq:cond2}
\end{equation}
The same holds true for all other subclusters, i.e.\ inequalities~\eqref{eq:cond2} must be obeyed for all $j=1,\ldots,s$. This
gives the second set of conditions for the composition $\{m_1,\ldots,m_s\}$ to occur.

For completeness, let us derive these conditions on the forces in a more formal manner by considering the equations of motions of particles 
in a subcluster of size $m$. For simplicity, we here label the particle coordinates in the subcluster
as  $x_1\, \ldots, x_m$. The equations are:
\vspace{-1ex}
\begin{subequations}
\label{eq:motions}
\begin{align}
\dot x_1&=\mu(F^{\rm ext}_1-F^{\rm int}_{1,2})\,,
\label{eq:motions-a}\\
\dot x_i&=\mu(F^{\rm ext}_i+F^{\rm int}_{i-1,i}-F^{\rm int}_{i,i+1})\,,\hspace{1em} i=2,\ldots,m\!-\!1\,,
\label{eq:motions-b}\\
\dot x_m&=\mu(F^{\rm ext}_m+F^{\rm int}_{m-1,m}) \,.\label{eq:motions-c}
\end{align}
\end{subequations}
All particles in the subcluster have the same velocity,
\begin{equation}
\dot x_1=\ldots=\dot x_m=\mu \bar F^{\rm ext}\,,
\label{eq:dotx-equal}
\end{equation}
where $\bar F^{\rm ext}=\sum_{i=1}^m F^{\rm ext}_i/m$ is the mean external force. 
Unless the external forces $F^{\rm ext}_i$ acting on the particles in the subcluster are all equal,
the interaction forces $F^{\rm int}_{i-1,i}$ between the particles must enforce these conditions.

In fact, Eqs.~\eqref{eq:dotx-equal} 
constitute $m$ linear equations for determining the $(m\!-\!1)$ interaction forces 
$F^{\rm int}_{1,2}, F^{\rm int}_{2,3},\ldots, F^{\rm int}_{m-1,m}$.
Only $(m\!-\!1)$ of these equations are independent, since $\sum_{i=1}^m \dot x_i$ does not depend on the interaction forces. The solution of the system of linear equations gives
\begin{equation}
F^{\rm int}_{i,i+1}=\frac{m\!-\!i}{m}\sum_{j=1}^i F^{\rm ext}_j-\frac{i}{m}\sum_{j=i+1}^m F^{\rm ext}_j\,.
\end{equation}
for $i=1,\ldots,m-1$.
These interaction forces have to be non-negative, yielding the $(m\!-\!1)$ conditions
in Eq.~\eqref{eq:cond2} for the subcluster to move as a whole (with $m_j=m$).

Moreover, each subcluster must become separated from its neighboring subclusters. 
This gives the set of conditions~\eqref{eq:cond1}.

\subsection{Algorithm}
\label{subsec:algorithm-hard-rods}
We consider the positions $x_i$ of the $N$ rods to be ordered as
\begin{equation}
x_1< x_2<\ldots< x_N\,,
\end{equation}
where $(x_{i+1}-x_i)\ge\sigma$. 
In the absence of the hardcore interaction, the particle positions could be propagated as
\begin{equation}
x_i(t+\Delta t)=x_i(t)+\mu f(x_i)\Delta t + \sqrt{2D\Delta t}\,\mathcal{N}_i\,,
\label{eq:Euler-propagation}
\end{equation}
when applying the Euler scheme with time step $\Delta t$ for solving Eqs.~\eqref{eq:langevin}.
Here $\mathcal{N}_i$ are independent random numbers drawn from a Gaussian distribution
with zero mean and unit variance (standard normal distribution). Accordingly, the total external forces 
\eqref{eq:fexttot} at time $t$ are approximated by
\begin{equation}
F^{\rm ext}_i(t)=\frac{x_i(t+\Delta t)-x_i(t)}{\mu\,\Delta t}=
f(x_i(t))+\frac{1}{\mu}\sqrt{\frac{2D}{\Delta t}}\,\mathcal{N}_i\,.
\label{eq:force_finite_dt}
\end{equation}

Knowing these forces, the cluster analysis is carried out. 
For each cluster of touching particles, we identify the decomposition into subclusters according to the condition \eqref{eq:cond1}. Specifically, for a $k$-cluster with $k>1$, we consider the 
fragmentation into each possible pair of subclusters ($j$, $k-j$), $j =1,\ldots,k-1$.
For these subclusters, we calculate the respective mean external forces 
$\bar F^{\rm ext}_j$ and $\bar F^{\rm ext}_{k-j}$, 
using Eq.~\eqref{eq:force_finite_dt} for the force exerted on each individual particle of the subcluster.
We split the $k$-cluster into that pair of subclusters, for which the difference 
$(\bar F^{\rm ext}_{k-j} - \bar F^{\rm ext}_j )> 0$ is the largest. 
Then we repeat this procedure for the two subclusters, and again for the pair of subclusters resulting 
from the two subclusters and so on. The procedure stops, if no further splitting into subclusters occurs.

After this step, it is known which particles form clusters that will
move with the same velocity at the beginning of the time interval $[t,t+\Delta t[$. 
For particles $i=j,\ldots,j+n$ forming such cluster with $n$ particles, the
velocities according to Eq.~\eqref{eq:dotx-equal} are
\begin{equation}
v_i(t)=\frac{\mu}{n}\sum_{k=j}^{j+n} F^{\rm ext}_k(t)\,,
\end{equation}
with the $F^{\rm ext}_k(t)$  from Eq.~\eqref{eq:force_finite_dt}.

Knowing the velocities, all particle positions are propagated within the time interval $[t,t+\Delta t[$ in an event-driven procedure from a collision at time $t_{\rm c}$ to a next collision
at time $t_{\rm c}'$. 

Initially we set $t_{\rm c}=t$. A possible 
collision can occur between neighboring rods, if their velocities satisfy
\begin{equation}
v_i(t_{\rm c})>v_{i+1}(t_{\rm c})\,.
\label{eq:collision-condition}
\end{equation}
The time of this possible collision is
\begin{equation}
t_{i,i+1}=t_{\rm c}+\frac{x_{i+1}(t_{\rm c})-x_i(t_{\rm c})-\sigma}{v_i(t_{\rm c})-v_{i+1}(t_{\rm c})}\,.
\end{equation}
The next collision realized is the one occuring at the smallest time in the set $\{t_{i,i+1}\}$, i.e.\
\begin{equation}
t_{\rm c}'=\min_i\{t_{i,i+1}\}\,.
\end{equation}

If $t_{\rm c}'<(t+\Delta t)$, we determine the particle $i_{\rm c}$ taking part in this collision:
\begin{equation}
i_{\rm c}=\argmin_i\{t_{i,i+1}\}\,.
\end{equation}
This collision leads to the merging of two clusters, where in the first cluster the rod 
$i_{\rm c}$ is the rightmost one and in the second cluster the rod $(i_{\rm c}+1)$ is the leftmost one.
As a consequence, after time $t_{\rm c}'$ the particles in the merged cluster move together and must 
have the same velocity, which is determined by the average external force, 
see Eq.~\eqref{eq:dotx-equal}. This implies that
the velocities of the particles in the merging clusters become equal and are given by the weighted average of the velocities $v_{i_{\rm c}}(t_{\rm c})$ and $v_{i_{\rm c}+1}(t_{\rm c})$.
If the first and second cluster contain $m$ rods and $n$ rods, respectively (merger of an $m$- with an $n$-cluster), the velocities $v_j$ of the rods $j=i_{\rm c}-m+1,i_{\rm c}-m+2,\ldots,i_{\rm c}+n$
at time $t_{\rm c}'$ become
\begin{equation}
v_j(t_{\rm c}')=\frac{m\,v_{i_{\rm c}}(t_{\rm c})+n\,v_{i_{\rm c}+1}(t_{\rm c})}{m+n}\,.
\end{equation}
This change of velocities means that the two clusters undergo a perfectly inelastic collision.
The velocities of all other rods are kept, i.e.\ $v_i(t_{\rm c}')=v_i(t_{\rm c})$ (no updating).

%%%%%%%%%%%%%%%%%%%%%%%%%%%%%%%%%%%%%%%%%%%%%%%%%%
\begin{figure}[b!]	
\centering
\includegraphics[width=\columnwidth]{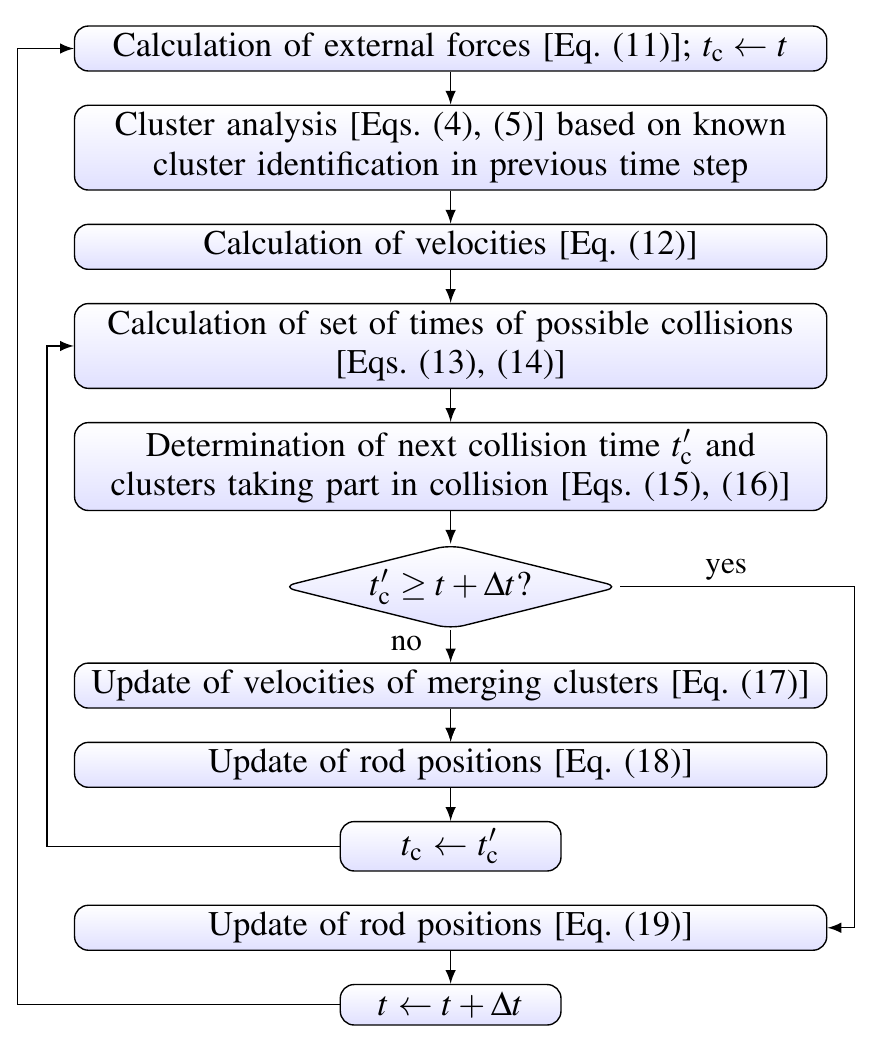}
\caption{Flow diagram of the BCD simulation algorithm \cite{Antonov/Schweers:2022}.}
\label{fig:flow-diagram} 
\end{figure}
%%%%%%%%%%%%%%%%%%%%%%%%%%%%%%%%%%%%%%%%%%%%%%%%%%

To keep the implementation simple, the positions $x_i$ of all rods are updated
at time $t_{\rm c}'$,
\begin{equation}
x_i(t_{\rm c}')=x_i(t_{\rm c})+v_i(t_{\rm c})(t_{\rm c}'-t_{\rm c})\,,\hspace{1em} i=1,\ldots,N\,.
\label{eq:position-update}
\end{equation}

This event-driven procedure given by Eqs.~\eqref{eq:collision-condition}-\eqref{eq:position-update} is repeated until $t_{\rm c}'\ge t+\Delta t$. Then
the positions of all rods $i=1,\ldots,N$ at time $t+\Delta t$ are calculated using
\begin{equation}
x_i(t+\Delta t)=x_i(t_{\rm c})+v_i(t_{\rm c})(t+\Delta t-t_{\rm c})\,.
\label{eq:position-update-final}
\end{equation}

Thereafter, updated external forces $F_i^{\rm ext}(t+\Delta t)$ at time $t+\Delta t$ are calculated, the
cluster analysis is performed again, and the event-driven procedure is carried out for the next time step.

We note that one could also update the external forces and perform the cluster analysis after each collision. This would lead to an implementation of the BCD algorithm with a variable time step. However, the use of a fixed time step becomes more efficient in particular in dense systems with high collision rate because the external forces do not need to be updated during a time interval $\Delta t$.

A flow diagram illustrating the algorithm is given in Fig.~\ref{fig:flow-diagram}.
A \Cpp{} code implementing this BCD algorithm, which uses vector manipulation
procedures from Refs.~\cite{Sanderson/Curtin:2016, Sanderson/Curtin:2018},
is available on \cite{Antonov/Schweers:2022}.

\subsection{Validation: Comparison with exact\\ analytical results}
\label{subsec:validation-hardcore-interaction}
For testing our algorithm, we compare simulation results for density profiles $\varrho(x)$ and two-particle densities 
$\varrho_2(x,x+\sigma)$
at contact with analytical results derived from the exact density functional \cite{Percus:1976}
\begin{subequations}
\begin{align}
&\Omega[\varrho(x)]\label{eq:percus_hc}\\
&=\int \dd x\, \varrho(x) 
\biggl\{U(x)-\mu_{\rms ch} -k_{\rms{B}} T \left[ 1-\ln\left(\frac{\varrho(x)}{1-\eta(x)} \right) \right]\biggr\}\nonumber
\end{align}    
of hard rods in equilibrium, where $U(x)$ is an external potential, 
$\mu_{\rms ch} $ is the chemical potential, and
\begin{equation}
\eta(x) = \int\limits_{x-\sigma}^{x}\dd y\,\varrho(y).
\label{eq:eta}
\end{equation}
\end{subequations}

%%%%%%%%%%%%%%%%%%%%%%%%%%%%%%%%%%%%%%%%%%%%%%%%%%
\begin{figure}[b!]	
\centering
\includegraphics[width=\columnwidth]{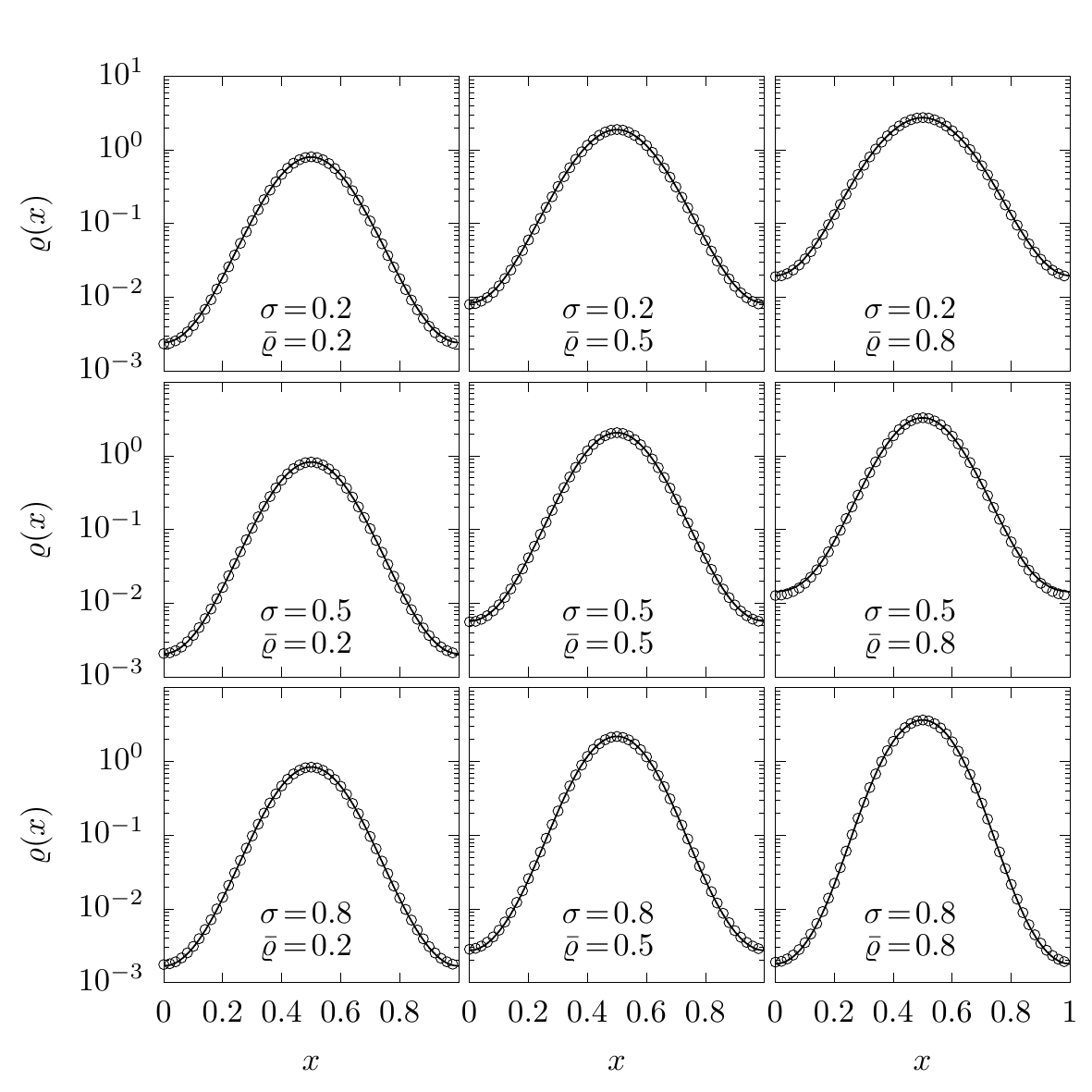}
\caption{Equilibrium density profiles of hard rods in the cosine potential \eqref{eq:cosine-potential}
obtained from BCD simulations (symbols) and from calculations based on the exact density 
functional [Eq.~\eqref{eq:percus_hc}]. The panels show results for various filling factors $\bar\varrho$ 
and rod lengths $\sigma$. The potential amplitude is $U_0=6\,k_{\rms B} T$.}
\label{fig:hard_rods_density} 
\end{figure}
%%%%%%%%%%%%%%%%%%%%%%%%%%%%%%%%%%%%%%%%%%%%%%%%%%

%%%%%%%%%%%%%%%%%%%%%%%%%%%%%%%%%%%%%%%%%%%%%%%%%%
\begin{figure}[t!]	
\centering
\includegraphics[width=\columnwidth]{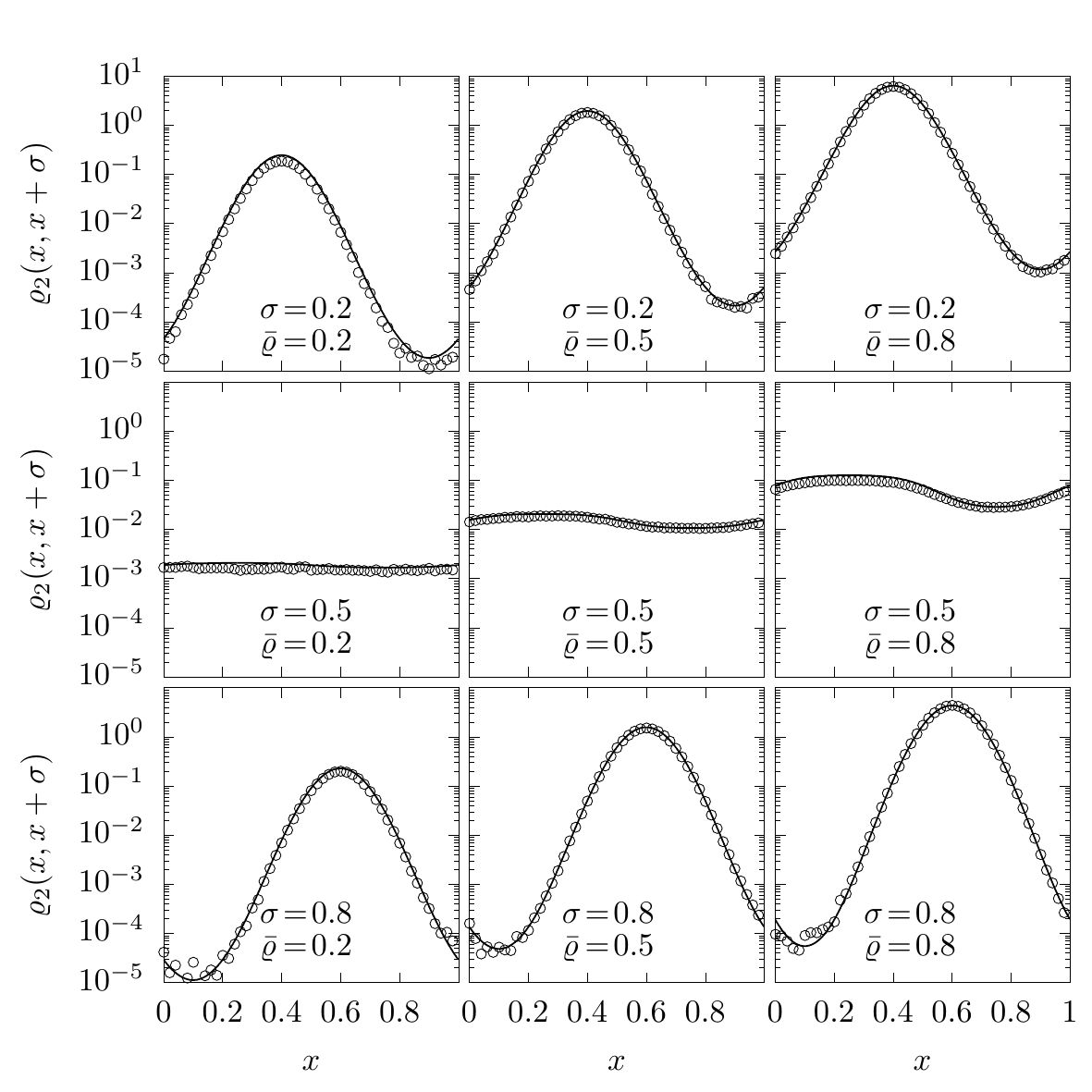}
\caption{Two-particle density at contact in equilibrium for hard rods in the cosine potential \eqref{eq:cosine-potential}
obtained from BCD simulations (symbols) and from calculations based on the 
exact density functional [Eq.~\eqref{eq:percus_hc}]. The panels show results for various filling factors $\bar\varrho$ and rod lengths $\sigma$. The potential amplitude is $U_0=6\,k_{\rms B} T$.}
\label{fig:hard_rods_contact_density}
\end{figure}
%%%%%%%%%%%%%%%%%%%%%%%%%%%%%%%%%%%%%%%%%%%%%%%%%%

We  choose the periodic potential
\begin{equation}
U(x)=\frac{U_0}{2}\cos\left(\frac{2\pi x}{\lambda}\right)\,,
\label{eq:cosine-potential}
\end{equation}
for which the integration limit in Eq.~\eqref{eq:percus_hc} can be taken from zero to the wavelength $\lambda$.

The equilibrium density profiles follow by minimization of the density functional $\Omega[\varrho(x)]$. Numerically, we
perform this minimization by evolving an initial profile using the 
dynamical density functional theory (DDFT) \cite{Marconi/Tarazona:1999, teVrugt/etal:2020} until the stationary equilibrium profile is obtained.
As initial profile, we used the Boltzmann one $\varrho(x)\propto\exp[-U(x)/k_{\rms B}T]$ of noninteracting particles with a normalization
\begin{equation}
\bar\varrho=\frac{1}{\lambda}\int_0^\lambda\dd x \varrho(x)=\frac{N}{L}\,,
\end{equation}
where the system length $L$ is a multiple of $\lambda$. The coverage (``volume fraction'') is 
equal to $\bar\varrho \sigma=N\sigma/L$.

The two-particle density at contact is \cite{Percus:1976,Marconi/Tarazona:2000}:
\begin{equation}
\varrho_2(x,x+\sigma) = \frac{\varrho(x)\varrho(x+\sigma)}{1-\eta(x+\sigma)}\,.
\end{equation}

To determine the density profile from our simulations, we discretize our periodicity interval into $M_{\rm bin}$ bins 
$[x_{\alpha},x_{\alpha}+\Delta_x[$ with 
\begin{equation}
x_{\alpha} = \alpha \Delta_x\,,\hspace{1em} \Delta_x = 1/M_{\rm bin}\,.
\end{equation}
After each time step $\Delta t$ in the simulation, we increase a counter $p_{\alpha}$ by the number of particles in bin $\alpha$. 
We thus obtain
\begin{equation}
 \varrho(x_\alpha + \Delta_x/2)=\frac{p_{\alpha}}{M_{\rm smp}\Delta_x}
 +\mathcal{O}(\Delta_x,M_{\rm smp}^{-1})\,,
 \label{eq:rho-numerical}
\end{equation}
where $M_{\rm smp}$ is the number of time intervals used in the sampling.
Equation~\eqref{eq:rho-numerical} allows one 
to calculate the density at position $x=x_\alpha+\Delta_x/2$ with controlled accuracy.
	
For determining the two-particle density at contact, we check after each time step whether one	particle 
with coordinate $x$ falls into the bin $\alpha$ and another
particle coordinate falls into the interval $]x + \sigma,x+\sigma + \Delta^{\prime}]$,
where $\Delta^{\prime} \ll 1$ is a small length. If this is the case, we increase
a counter $q_{\alpha}$ for the bin $\alpha$ by one. 
We thus obtain
\begin{align}
\varrho_2(x_\alpha + \Delta_x/2,x_\alpha + \Delta_x/2+\sigma)=&\\
&\hspace{-5em}\frac{q_{\alpha}}{M_{\rm smp}\Delta_x\Delta'}+\mathcal{O}(\Delta_x,\Delta',M_{\rm smp}^{-1})\nonumber
 \label{eq:rho2-numerical}
\end{align}
for calculating the two-particle density at contact if the left particle is at position $x=x_\alpha+\Delta_x/2$. 

In the following we take $\lambda$ as length unit, $\lambda^2/D$ as time unit,
and $k_{\rms B}T$ as energy unit ($\lambda=1$, $D=1$, $k_{\rms B}T=1$).
Furthermore we must set $D=k_{\rms B}T\mu$ for simulating equilibrium properties, i.e.\ $\mu=1$ in our units.
We consider a high amplitude $U_0=6$ of the external potential.
In the simulations, we used the system length $L =100$ with periodic boundary conditions, 
and chose $\Delta_x=\Delta^{\prime}=10^{-3}$
for the discretization lengths. The time step is reduced with increasing particle density $\bar\varrho$ to properly resolve collisions between clusters. For the largest simulated $\bar\varrho=0.8$ here, we chose $\Delta t=10^{-7}$.

In Figs.~\ref{fig:hard_rods_density} and \ref{fig:hard_rods_contact_density}
we compare simulated equilibrium density profiles $\varrho(x)$ with the theoretical profiles derived from the Percus functional. 
These figures show an excellent agreement between theory and simulation for various representative values of rod lengths $\sigma$ and 
mean particle densities $\bar\varrho$.

\section{Additional interactions}
\label{sec:additional-interactions}
If there are additional interaction forces between the hard rods, it is straightforward to include them in the numerical treatment as long as the interactions are continuous functions of the particle coordinates. 
We focus here on pair interactions and denote by $f^{\rm int}(x_i,x_j)$ the force of rod $j$ exerted on rod $i$. Such interactions can include van der Waals, electrostatic, magnetic, steric and/or depletion forces \cite{Lewis:2000}.
The Langevin equations for overdamped Brownian motion of the rods then are
\begin{equation}
\frac{\dd x_i}{\dd t}=\mu f(x_i)+\mu \sum_{j=1}^N f^{\rm int}(x_i,x_j)+\sqrt{2D}\,\xi_i(t)\,,
\label{eq:langevin-with-interactions}
\end{equation}
where we set $f^{\rm int}(x_i,x_i)=0$. 

When applying the algorithm discussed in Sec.~\ref{subsec:algorithm-hard-rods}, the only change 
is that the external forces in Eq.~\eqref{eq:force_finite_dt} need to be replaced by
\begin{equation}
F^{\rm ext}_i(t)=
f(x_i(t))+\sum_{j=1}^N f^{\rm int}(x_i(t),x_j(t))
+\frac{1}{\mu}\sqrt{\frac{2D}{\Delta t}}\,\mathcal{N}_i\,.
\label{eq:force_finite_dt_with_interaction}
\end{equation}

The prominent potential for sticky hard spheres contains a $\delta$-singularity and
accordingly does not give a continuous interaction force between the rods. It therefore requires a special treatment.

\subsection{Treatment of Baxter's sticky hard-sphere potential}
\label{subsec:sticky-core-potential}
Baxter introduced his model of sticky hard spheres to capture major characteristics of real interactions, which
exhibit a repulsive core and attractive part. If the attractive part is a short-range surface adhesion between particles, the pair interaction potential $V_{\rm int}(r)$ may be modeled by including a 
$\delta$-function in the corresponding Boltzmann factor,
\begin{equation}
\exp[-V_{\rm int}(r)/k_{\rms B} T] = \Theta(r-\sigma)+\gamma\delta(r-\sigma)\,.
\end{equation}
Here, $\Theta(.)$ is the Heaviside step function, $k_{\rms B} T$ is the thermal energy, and $r$ is the distance between particle positions. The
parameter $\gamma$ is a measure for the strength of the adhesive interaction.
This yields the potential \cite{Baxter:1968,Percus:1982}
\begin{equation}
V_{\rm int}(r) = \left\{\begin{array}{lc}
\infty, & r < \sigma, \\[0.5ex]
-k_{\rms B}T \ln[1 + \gamma\delta(r - \sigma)], & r \ge \sigma\,.
\end{array}\right.
\label{eq:sticky-core-potential}
\end{equation}
In the recent literature, particles with this type of interaction are often referred to as adhesive hard spheres.

Baxter's model has been applied as an approximation to understand properties of Lennard-Jones fluids \cite{Watts:1969, Gonzalez-Calderon:2019}. Interestingly, when increasing the exponents in the Lennard-Jones potential, possible cluster formations approach the ones of Baxter's model \cite{Trombach/etal:2018}. 
It was considered also to uncover effects of adhesive interaction on percolation behavior \cite{Chiew/Glandt:1983, Seaton/Glandt:1987, Seaton/Glandt:1987_2, Torquato/etal:1988, Kim/etal:2014}. 
Further applications include analysis of colloid sedimentation on liquid-solid  \cite{Piazza:2014} and fluid-fluid interfaces \cite{Balazs/etal:2020}, different types of sequential adsorption \cite{Talbot/etal:2000}, gelation \cite{Miller/Frenkel:2004_1, Zaccarelli:2007, Richard/etal:2018, Wang/Swan:2019} and gel elasticity \cite{Whitaker/etal:2019},
aggregation of biomolecules \cite{Assenza/Mezzenga:2019, Smith/etal:2020}, 
crystallization of macromolecules \cite{Rosenbaum/etal:1996}, 
self-assembly of inactive \cite{Genix/Oberdisse:2018} and active \cite{Schwarz-Linek/etal:2012, dePirey/etal:2019} particles, detachment dynamics of colloidal spheres \cite{Bergenholtz:2018}, viscosity of adhesive hard sphere dispersions \cite{Wang/etal:2019, vonBuelow/etal:2019}, and charge regulation of colloids inside electrolyte solutions \cite{Bakhshandeh/etal:2020}. 
Another important application pertains to the analysis of scattering experiments on colloids \cite{Pedersen:1997, Svergun:2003} and nanoparticle mixtures \cite{Li/etal:2016, Motokawa/etal:2019, Ko/etal:2021, Heil/Jayaraman:2021, Jeffries/etal:2021, Heil/etal:2022}.  The model has been used also to describe
dynamics in glassy states and liquid-glass phase transitions \cite{Bergenholtz/Fuchs:1999, Dawson/etal:2000, Parisi/Zamponi:2010, Fullerton/Berthier:2020}. The model has been extended to binary mixtures of particles with different strength of Baxter's interactions \cite{Baxter:1970, Jamnik:2008} and binary mixtures of different particles sizes \cite{Opdam/etal:2021}. Under certain conditions (size disparity and high concentrations), such mixtures can undergo a fluid-fluid phase separation \cite{Kobayashi/etal:2021}. Furthermore, models based on Baxter's model have been suggested for mixtures of colloids and telechelic polymers \cite{Zhang/etal:2019, Amani/etal:2021} and for protein-protein interactions between proteins of different size \cite{Bye/Curtis:2019}. 

In order to deal with the $\delta$-singularity in Eq.~\eqref{eq:sticky-core-potential}, we use a power-law representation 
$\delta_\epsilon(r)$ 
of the $\delta$-function:
\begin{equation}
\delta_\epsilon(r) =\frac{\expo+1}{\epsilon^{\expo+1}}(\epsilon-r)^{\expo}
[\Theta(r)-\Theta(r-\epsilon)]\,,
\label{eq:deltaeps}
\end{equation}
where $\expo>1$. For $\expo\le1$, the force would no longer be continuous at $r=\sigma+\epsilon$.
The virtual interaction range $\epsilon>0$ needs to be much smaller than the average distance between neighboring particles. For $\epsilon\to0$, $\delta_\epsilon(r) \to\delta(r)$.

In Baxter's original work \cite{Baxter:1968}, the $\delta$-singularity in Eq.~\eqref{eq:sticky-core-potential}
was represented by a rectangular well, corresponding to $\expo=0$ in Eq.~\eqref{eq:deltaeps}.
This was particularly suitable for the solution of the Percus-Yevick equations. 
However, a rectangular potential well would 
yield $\delta$-singularities in the forces and is thus not a useful representation for simulation of Brownian dynamics based on Langevin equations. 

By inserting this representation in Eq.~\eqref{eq:sticky-core-potential}, we obtain the interaction force
\begin{equation}
f_p^{\rm int}(r) = \left\{\begin{array}{ll}
\displaystyle -k_{\rms B}T  \frac{\gamma \expo (\epsilon+\sigma-r)^{\expo-1}}{\frac{\epsilon^{\expo+1}}{\expo+1} + \gamma(\epsilon+\sigma-r)^{\expo}},& \sigma\leq r<\sigma+\epsilon\,,\\[1ex]
0,& r \geq \sigma + \epsilon\,.
\end{array}\right.
\label{eq:fp}
\end{equation}
Because of the constant one introduced in the argument of the logarithmic function 
in Eq.~\eqref{eq:sticky-core-potential},
this interaction force is continuous at $r=\sigma+\epsilon$.

While $\gamma$ is a physical parameter characterizing the strength of the adhesive interaction, 
$\expo$ and $\epsilon$ are auxiliary parameters defining the representation of the $\delta$-function in
Eq.~\eqref{eq:sticky-core-potential}. What is a good choice of these auxiliary parameters? 

The parameter $\epsilon$ gives a virtual range of the attractive contact interaction and should be 
much smaller than all relevant length scales in the system like the particle size, mean distance between particles, and characteristic lengths of the external force field, as, e.g., the wavelength in a periodic field.
At the same time, $\epsilon$ should not be taken too small in order
to resolve the virtual interaction range, when particles come close to each other and are driven by 
$f_p^{\rm int}$ towards each other. This can be accounted for by imposing the condition
\begin{equation}
\mu \max_r\{\abs{f_{\expo}^{\rm int}(r)}\}\Delta t \ll\epsilon\,,
\label{eq:ConditionInteraction}
\end{equation}
which means that the maximum displacement caused by the adhesive 
interaction in a time step $\Delta t$ is much smaller than $\epsilon$.

\begin{figure}[t!]	
\centering
\includegraphics[scale=1]{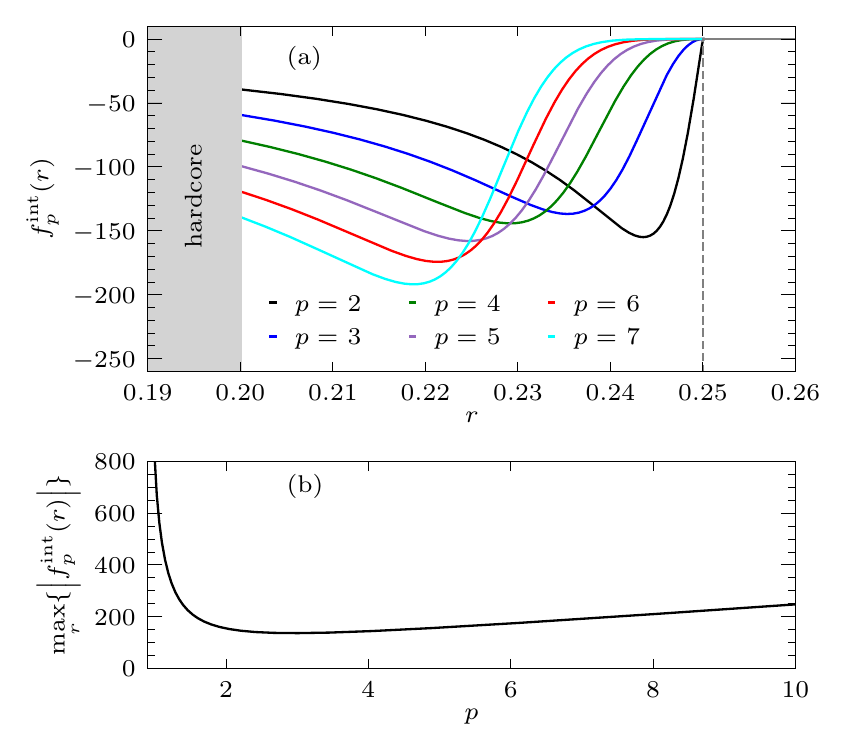}
\caption{(a) Interaction force $f_{\expo}^{\rm int}(r)$ [Eq.~\eqref{eq:fp}] 
exerted on a particle at distance $r$ by a particle at the origin
for various $\expo$, $\epsilon= 0.05$, 
$\gamma=1$, and $\sigma=0.2$. The hardcore regime $r\le\sigma$ is marked by the shaded area. The vertical dashed lines indicates the end of the virtual interaction interval at $\sigma+\epsilon=0.25$. The interaction forces are continuous at this point and become zero for larger $r$, as indicated by the grey horizontal line. (b) Maximum of
$\abs{f_{\expo}^{\rm int}(r)}$ as a function of the exponent in the power-law representation $\delta_\epsilon(r)$ in Eq.~\eqref{eq:deltaeps} for the same parameters as in (a).}
\label{fig:Poly}
\end{figure}

In Fig.~\ref{fig:Poly}(a), we plot $f_{\expo}^{\rm int}(r)$ for $\epsilon=0.05$ and various $\expo$. 
For this $\epsilon$, Fig.~\ref{fig:Poly}(b) shows that $\max_r\{\abs{f_{\expo}^{\rm int}(r)}\}$ is smallest for $\expo\cong2.95$.
As $\epsilon$ should be as small as possible in Eq.~\eqref{eq:ConditionInteraction}, this value for $\expo$ is preferable.
For this reasoning, we used that $\max_r\{\abs{f_{\expo}^{\rm int}(r)}\}$ is a weakly varying function of $\epsilon$.

%%%%%%%%%%%%%%%%%%%%%%%%%%%%%%%%%%%%%%%%%%%%%%%%%%
%%%%%%%%%%%%%%%%%%%%%%%%%%%%%%%%%%%%%%%%%%%%%%%%%%
	
\subsection{Validation: Comparison with exact\\ analytical results}
\label{subsec:validation-contact-interactions}
Similarly as in Sec.~\ref{subsec:validation-hardcore-interaction}, we test our treatment of 
Baxter's potential in connection with the algorithm described in Sec.~\ref{subsec:algorithm-hard-rods} by comparison with
exact analytical results for equilibrium density profiles. 
These profiles are generated for the periodic potential in Eq.~\eqref{eq:cosine-potential}
by applying the dynamic density functional theory, where now the
exact functional derivative of the density functional for hard rods with contact interaction \cite{Percus:1982} is used:
\begin{subequations}
\label{eq:dOm-contact-interaction}
\begin{align}
&\frac{\delta\Omega[\varrho(x)]}{\delta \varrho(x)} = \ln\left(\frac{K(x+\sigma)\hat{K}(x-\sigma)}{\varrho(x)}\right) + \frac{U(x) - \mu_{\rms ch}}{k_{\rms B}T}\nonumber\\
&+ \frac{1}{2}\ln\left(1 - \eta(x)\right)\left(1 - \eta(x+\sigma)\right)-\frac{\sigma}{2\gamma} \label{eq:stickyDDFT}\\
&+\hspace{-0.3em}\int\limits_{x-\sigma}^{x}\frac{\dd y}{2\gamma}\,
\sqrt{1\!+\!2\gamma\frac{\varrho(y\!+\!\sigma)\!+\!\varrho(y)}{1-\eta(y\!+\!\sigma)}
+\gamma^2\!\left[\frac{\varrho(y\!+\!\sigma)-\varrho(y)}{1-\eta(y\!+\!\sigma)}\right]^2}\nonumber
\end{align}
with
\begin{align}
&\hat{K}(x)=\frac{1}{2\gamma}\Biggl[-\left(1-\gamma\frac{\varrho(x+\sigma)-\varrho(x)}{1-\eta(x+\sigma)}\right)\\
&\hspace{1em}{}\left.+\sqrt{\left(1 -\gamma\frac{\varrho(x+\sigma)-\varrho(x)}{1 - \eta(x+\sigma)}\right)^2 
+ 4\gamma\frac{\varrho(x+\sigma)}{1-\eta(x+\sigma)}}\,\right]\nonumber
\end{align}
and
\begin{equation}
K(x) = \hat{K}(x-\sigma)-\frac{\varrho(x)-\varrho(x-\sigma)}{1-\eta(x)}\,.
\label{eq:K(x)}
\end{equation}
\end{subequations}
The solution of the structure equation $\delta\Omega[\varrho(x)]/\delta \varrho(x)=0$ gives the equilibrium density profile. As in the case of hard rods without contact
interaction, we generated this equilibrium profile by evolving the corresponding DDFT equations into the stationary long-time limit. 
The computational time for reaching this long-time limit is much longer than for the hard rods without contact interaction, because 
of the more complex mathematical structure of the functional derivative  $\delta\Omega[\varrho(x)]/\delta \varrho(x)$ in Eqs.~\eqref{eq:stickyDDFT}-\eqref{eq:K(x)}.
For obtaining results in reasonable computing time,
we were limited to use a spatial discretization length $\Delta x=10^{-3}$ 
(in units of $\lambda$) \footnote{The computational time for generating the stationary profile in the DDFT increases as $\Delta x^{-2}$. Our choice $\Delta x=10^{-3}$
allowed us to obtain the equilibrium profile in a computational time of about one day with an Intel Core i5-7600 CPU 3.50GHz
processor. For a spatial resolution $\Delta x=10^{-4}$ the computational time would be 100 days.}. As we discuss below,
this limitation causes small systematic deviations at high coverages $\bar\varrho\sigma$ in spatial regions of low particle densities.

In the simulations, all parameters are chosen as for the hard rods without contact interaction considered in 
Sec.~\ref{subsec:validation-hardcore-interaction}. To represent the contact interaction we use the integer exponent $\expo=3$ in Eq.~\eqref{eq:deltaeps} and set $\epsilon=0.05$.

\begin{figure}[t!]	
\centering
\includegraphics[width=\columnwidth]{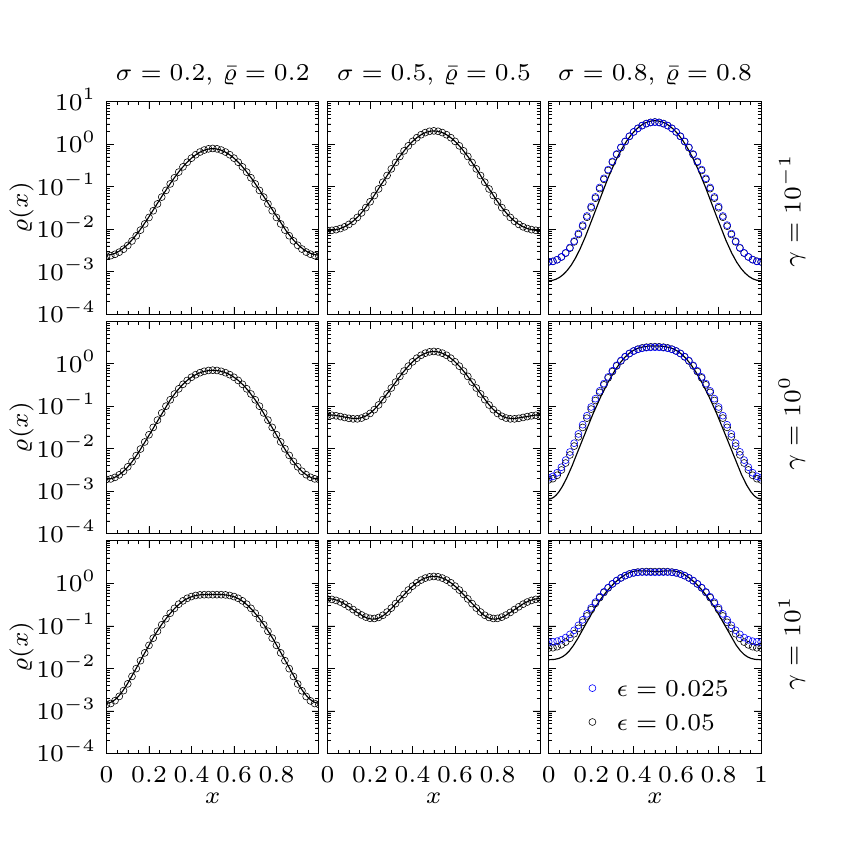}
\caption{Equilibrium density profiles of sticky hard rods 
in the cosine potential \eqref{eq:cosine-potential}
obtained from BCD simulations (symbols) and from numerical solution of the DFT structure equations with the functional derivative from
Eqs.~\eqref{eq:dOm-contact-interaction} (solid lines). 
The panels in each column refer to equal values of the mean particle density $\bar\varrho$ and rod length $\sigma$ given at the top of the figure. The rows refers to equal values of the strength $\gamma$ of the contact interaction 
\eqref{eq:sticky-core-potential}. Its values are given on the right side of the figure.
The amplitude of the cosine potential is $U_0=6\,k_{\rms B} T$. In the graphs for $\sigma=0.8$ and 
$\bar\varrho=0.8$ (right column), density profiles are shown for two values of the virtual interaction range $\epsilon$.}
\label{fig:DensityProfilesSticky}
\end{figure}

In Fig.~\ref{fig:DensityProfilesSticky}, simulated equilibrium density profiles $\varrho(x)$ are compared
with theoretical profiles calculated from Eqs.~\eqref{eq:dOm-contact-interaction} for
three strengths $\gamma$ of the contact interaction and various representative rod lengths $\sigma$
and mean particle densities $\bar\varrho=0.2$, 0.5, and 0.8. For $\bar\varrho=0.2$ and 0.5, the simulated 
data show an excellent agreement with the analytical predictions. 

For $\bar\varrho=0.8$ (graphs in the right column in Fig.~\ref{fig:DensityProfilesSticky}), 
the agreement between simulations and the numerical solution of the structure equations is also good, but close to $x=0$ (or $x=1$)
small deviations are seen. Note that the density profiles in Fig.~\ref{fig:DensityProfilesSticky} are plotted on a logarithmic scale, i.e.\ 
the absolute deviation between the data obtained from the simulations and the numerical solution is very small.

In order to see whether the deviations are caused by our choice of $\epsilon$ or the limited accuracy of the numerical solution of the structure equations, we 
in addition carried out simulations for a smaller value $\epsilon=0.025$ (blue circles in the right column of Fig.~\ref{fig:DensityProfilesSticky}). 
For $\gamma=0.1$ and $\gamma=1$, the simulated results for $\epsilon=0.05$ and $\epsilon=0.025$ overlap, and for $\gamma=10$ 
the results are almost converged
with respect to a decrease of $\epsilon$.
This indicates that the simulated data represent the true profiles and that 
the numerical solution of the structure equation is not perfectly accurate close to $x=0$. As discussed above, the accuracy of solving the structure equations
is limited by the spatial  resolution $\Delta x$.
	
%%%%%%%%%%%%%%%%%%%%%%%%%%%%%%%%%%%%%%%%%%%%%%%%%%
%%%%%%%%%%%%%%%%%%%%%%%%%%%%%%%%%%%%%%%%%%%%%%%%%%

\subsection{Time-dependent mean-square displacement}
\label{subsec:msd}
\begin{figure}[t!]	
\centering
\includegraphics[width=\columnwidth]{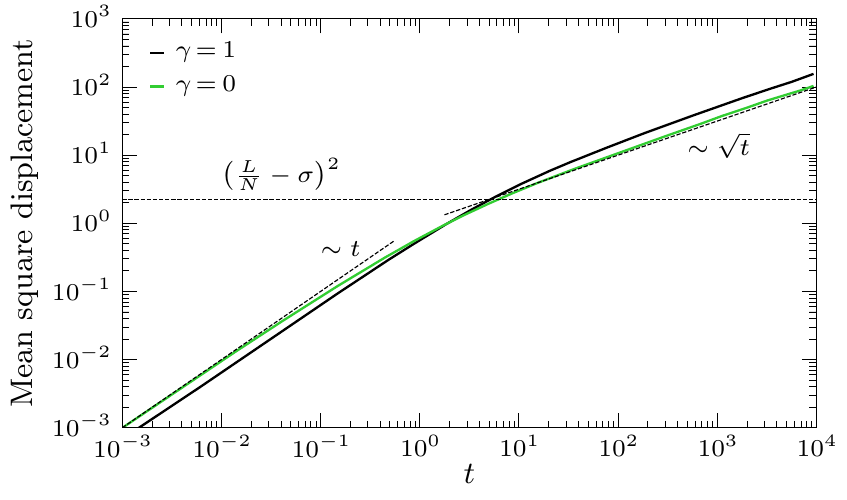}
\caption{Time-dependent mean-square displacement of a tagged hard rod for
a system with $U_0=0$ (no periodic potential) in the absence of a contact interaction ($\gamma=0$), and for sticky rods with $\gamma=1$. 
The  rod length is $\sigma=0.5$ and the mean density is $\bar\varrho=0.5$. The data were obtained
from simulations in a system of length $L=400$ with periodic boundary conditions. Dashed lines at short and long times indicate the asymptotic diffusive and subdiffusive behavior, and the dashed horizontal line marks the squared mean interparticle distance.}
\label{fig:msd}
\end{figure}

As an application of our method to a dynamical quantity, we show in Fig.~\ref{fig:msd} the time-dependent mean-square displacement (MSD)
of a tagged rod for a system with $U_0=0$ (no periodic potential) in the absence of a contact interaction ($\gamma=0$), and for sticky rods with $\gamma=1$. 
At short times, where the root of the MSD is smaller than the mean interparticle spacing, the dynamics must reflect that of independent rods, i.e.\ the MSD increases linearly with time. The short-time diffusion coefficient is smaller in the presence of contact interaction due to enhanced cluster formation.
When the MSD becomes comparable to the 
squared mean interparticle distance [$(L/N-\sigma)^2$, as indicated by the dashed horizontal line in Fig.~\ref{fig:msd}], collisions become relevant. The behavior then
crosses over to the well-known anomalous single-file diffusion, where the MSD grows as a square root of $t$. Interestingly, the subdif\-fusion is faster in the presence of contact interaction. A more detailed investigation of this effect and the change of behavior for $U_0>0$ will be presented elsewhere.

%%%%%%%%%%%%%%%%%%%%%%%%%%%%%%%%%%%%%%%%%%%%%%%%%%
%%%%%%%%%%%%%%%%%%%%%%%%%%%%%%%%%%%%%%%%%%%%%%%%%%

\section{Summary and Conclusion}
\label{sec:conclusion}
Overdamped Brownian motion of particles is ubiquitous in soft matter and biological systems and clustering of particles is widely observed in the dynamics observed at high densities. Here we have presented a new method to simulate Brownian dynamics of clusters formed by hardcore interacting particles in one dimension. This method can be applied to the motion in arbitrary external force fields and allows one to determine interaction forces between particles in contact. It relies on a fragmentation and merging of clusters based on the evaluation of mean external forces acting upon individual clusters. This includes the random forces exerted by the fluid environment. We proposed an algorithm to implement the fragmentation and merging processes in an event-driven scheme.

Particle interactions beyond hardcore can be straightforwardly included into the algorithm as long as the corresponding 
interaction forces are continuous functions of the particle coordinates. Other types of interactions need addition considerations. In particular, Baxter's sticky hard-sphere interaction for describing adhesive forces is an important case. We developed a method to tackle the $\delta$-singularity in this model which occurs when particles get in contact.

Our event-driven numerical method for treating cluster dynamics in collective overdamped Brownian motion has been 
successfully tested by comparison with theoretical results derived from exact density functionals, both for hardcore interacting systems and in the presence of the additional attractive contact interaction in the sticky hard-sphere model. 

In view of the existing methods for treating Brownian motion of hard spheres,
we consider our new method to fill a gap when it becomes necessary to model collective dynamics of particle clusters, as they can  manifest themselves, for example, as Brownian solitons \cite{Antonov/etal:2022a}. As for further developments of the method, a challenging task its extension to higher dimensions. This requires a refined procedure of cluster analysis and the cluster dynamics may be treated similarly as in the one-dimensional case by decomposing forces in the directions normal to the hard spheres at contact points and perpendicular to it. 
	
\section{Acknowledgements}
Financial support by the Czech Science Foundation (Project No.\ 20-24748J) and the Deutsche Forschungsgemeinschaft
(Project No.\ 432123484) is gratefully acknowledged.

%\bibliography{/Users/Maass/bibfiles/driven-systems}

\begin{thebibliography}{85}%
\makeatletter
\providecommand \@ifxundefined [1]{%
 \@ifx{#1\undefined}
}%
\providecommand \@ifnum [1]{%
 \ifnum #1\expandafter \@firstoftwo
 \else \expandafter \@secondoftwo
 \fi
}%
\providecommand \@ifx [1]{%
 \ifx #1\expandafter \@firstoftwo
 \else \expandafter \@secondoftwo
 \fi
}%
\providecommand \natexlab [1]{#1}%
\providecommand \enquote  [1]{``#1''}%
\providecommand \bibnamefont  [1]{#1}%
\providecommand \bibfnamefont [1]{#1}%
\providecommand \citenamefont [1]{#1}%
\providecommand \href@noop [0]{\@secondoftwo}%
\providecommand \href [0]{\begingroup \@sanitize@url \@href}%
\providecommand \@href[1]{\@@startlink{#1}\@@href}%
\providecommand \@@href[1]{\endgroup#1\@@endlink}%
\providecommand \@sanitize@url [0]{\catcode `\\12\catcode `\$12\catcode
  `\&12\catcode `\#12\catcode `\^12\catcode `\_12\catcode `\%12\relax}%
\providecommand \@@startlink[1]{}%
\providecommand \@@endlink[0]{}%
\providecommand \url  [0]{\begingroup\@sanitize@url \@url }%
\providecommand \@url [1]{\endgroup\@href {#1}{\urlprefix }}%
\providecommand \urlprefix  [0]{URL }%
\providecommand \Eprint [0]{\href }%
\providecommand \doibase [0]{https://doi.org/}%
\providecommand \selectlanguage [0]{\@gobble}%
\providecommand \bibinfo  [0]{\@secondoftwo}%
\providecommand \bibfield  [0]{\@secondoftwo}%
\providecommand \translation [1]{[#1]}%
\providecommand \BibitemOpen [0]{}%
\providecommand \bibitemStop [0]{}%
\providecommand \bibitemNoStop [0]{.\EOS\space}%
\providecommand \EOS [0]{\spacefactor3000\relax}%
\providecommand \BibitemShut  [1]{\csname bibitem#1\endcsname}%
\let\auto@bib@innerbib\@empty
%</preamble>
\bibitem [{\citenamefont {Rosenbluth}\ and\ \citenamefont
  {Rosenbluth}(1954)}]{Rosenbluth/Rosenbluth:1954}%
  \BibitemOpen
  \bibfield  {author} {\bibinfo {author} {\bibfnamefont {M.~N.}\ \bibnamefont
  {Rosenbluth}}\ and\ \bibinfo {author} {\bibfnamefont {A.~W.}\ \bibnamefont
  {Rosenbluth}},\ }\bibfield  {title} {\bibinfo {title} {Further results on
  {Monte Carlo} equations of state},\ }\href
  {https://doi.org/10.1063/1.1740207} {\bibfield  {journal} {\bibinfo
  {journal} {J. Chem. Phys.}\ }\textbf {\bibinfo {volume} {22}},\ \bibinfo
  {pages} {881} (\bibinfo {year} {1954})}\BibitemShut {NoStop}%
\bibitem [{\citenamefont {Alder}\ and\ \citenamefont
  {Wainwright}(1957)}]{Alder/Wainwright:1957}%
  \BibitemOpen
  \bibfield  {author} {\bibinfo {author} {\bibfnamefont {B.~J.}\ \bibnamefont
  {Alder}}\ and\ \bibinfo {author} {\bibfnamefont {T.~E.}\ \bibnamefont
  {Wainwright}},\ }\bibfield  {title} {\bibinfo {title} {Phase transition for a
  hard sphere system},\ }\href {https://doi.org/10.1063/1.1743957} {\bibfield
  {journal} {\bibinfo  {journal} {J. Chem. Phys.}\ }\textbf {\bibinfo {volume}
  {27}},\ \bibinfo {pages} {1208} (\bibinfo {year} {1957})}\BibitemShut
  {NoStop}%
\bibitem [{\citenamefont {Cichocki}\ and\ \citenamefont
  {Hinsen}(1990)}]{Cichocki/Hinsen:1990}%
  \BibitemOpen
  \bibfield  {author} {\bibinfo {author} {\bibfnamefont {B.}~\bibnamefont
  {Cichocki}}\ and\ \bibinfo {author} {\bibfnamefont {K.}~\bibnamefont
  {Hinsen}},\ }\bibfield  {title} {\bibinfo {title} {Dynamic computer
  simulation of concentrated hard sphere suspensions: I.\ {S}imulation
  technique and mean square displacement data},\ }\href
  {https://doi.org/https://doi.org/10.1016/0378-4371(90)90068-4} {\bibfield
  {journal} {\bibinfo  {journal} {Physica A}\ }\textbf {\bibinfo {volume}
  {166}},\ \bibinfo {pages} {473} (\bibinfo {year} {1990})}\BibitemShut
  {NoStop}%
\bibitem [{\citenamefont {Schaertl}\ and\ \citenamefont
  {Sillescu}(1994)}]{Schaertl/Sillescu:1994}%
  \BibitemOpen
  \bibfield  {author} {\bibinfo {author} {\bibfnamefont {W.}~\bibnamefont
  {Schaertl}}\ and\ \bibinfo {author} {\bibfnamefont {H.}~\bibnamefont
  {Sillescu}},\ }\bibfield  {title} {\bibinfo {title} {Brownian dynamics of
  polydisperse colloidal hard spheres: Equilibrium structures and random close
  packings},\ }\href {https://doi.org/10.1007/BF02183148} {\bibfield  {journal}
  {\bibinfo  {journal} {J. Stat. Phys.}\ }\textbf {\bibinfo {volume} {77}},\
  \bibinfo {pages} {1007} (\bibinfo {year} {1994})}\BibitemShut {NoStop}%
\bibitem [{\citenamefont {Jabbari-Farouji}\ and\ \citenamefont
  {Trizac}(2012)}]{Jabbari-Farouji/Trizac:2012}%
  \BibitemOpen
  \bibfield  {author} {\bibinfo {author} {\bibfnamefont {S.}~\bibnamefont
  {Jabbari-Farouji}}\ and\ \bibinfo {author} {\bibfnamefont {E.}~\bibnamefont
  {Trizac}},\ }\bibfield  {title} {\bibinfo {title} {Dynamic {M}onte {C}arlo
  simulations of anisotropic colloids},\ }\href
  {https://doi.org/10.1063/1.4737928} {\bibfield  {journal} {\bibinfo
  {journal} {J. Chem. Phys.}\ }\textbf {\bibinfo {volume} {137}},\ \bibinfo
  {pages} {054107} (\bibinfo {year} {2012})}\BibitemShut {NoStop}%
\bibitem [{\citenamefont {Corbett}\ \emph {et~al.}(2018)\citenamefont
  {Corbett}, \citenamefont {Cuetos}, \citenamefont {Dennison},\ and\
  \citenamefont {Patti}}]{Corbett/etal:2018}%
  \BibitemOpen
  \bibfield  {author} {\bibinfo {author} {\bibfnamefont {D.}~\bibnamefont
  {Corbett}}, \bibinfo {author} {\bibfnamefont {A.}~\bibnamefont {Cuetos}},
  \bibinfo {author} {\bibfnamefont {M.}~\bibnamefont {Dennison}},\ and\
  \bibinfo {author} {\bibfnamefont {A.}~\bibnamefont {Patti}},\ }\bibfield
  {title} {\bibinfo {title} {Dynamic {M}onte {C}arlo algorithm for
  out-of-equilibrium processes in colloidal dispersions},\ }\href
  {https://doi.org/10.1039/C8CP02415D} {\bibfield  {journal} {\bibinfo
  {journal} {Phys. Chem. Chem. Phys.}\ }\textbf {\bibinfo {volume} {20}},\
  \bibinfo {pages} {15118} (\bibinfo {year} {2018})}\BibitemShut {NoStop}%
\bibitem [{\citenamefont {Garc\'{\i}a~Daza}\ \emph {et~al.}(2020)\citenamefont
  {Garc\'{\i}a~Daza}, \citenamefont {Cuetos},\ and\ \citenamefont
  {Patti}}]{Daza/etal:2020}%
  \BibitemOpen
  \bibfield  {author} {\bibinfo {author} {\bibfnamefont {F.~A.}\ \bibnamefont
  {Garc\'{\i}a~Daza}}, \bibinfo {author} {\bibfnamefont {A.}~\bibnamefont
  {Cuetos}},\ and\ \bibinfo {author} {\bibfnamefont {A.}~\bibnamefont
  {Patti}},\ }\bibfield  {title} {\bibinfo {title} {Dynamic {M}onte {C}arlo
  simulations of inhomogeneous colloidal suspensions},\ }\href
  {https://doi.org/10.1103/PhysRevE.102.013302} {\bibfield  {journal} {\bibinfo
   {journal} {Phys. Rev. E}\ }\textbf {\bibinfo {volume} {102}},\ \bibinfo
  {pages} {013302} (\bibinfo {year} {2020})}\BibitemShut {NoStop}%
\bibitem [{\citenamefont {Hurtado}\ and\ \citenamefont
  {Garrido}(2020)}]{Hurtado/Garrido:2020}%
  \BibitemOpen
  \bibfield  {author} {\bibinfo {author} {\bibfnamefont {P.~I.}\ \bibnamefont
  {Hurtado}}\ and\ \bibinfo {author} {\bibfnamefont {P.~L.}\ \bibnamefont
  {Garrido}},\ }\bibfield  {title} {\bibinfo {title} {Simulations of transport
  in hard particle systems},\ }\href
  {https://doi.org/10.1007/s10955-019-02469-z} {\bibfield  {journal} {\bibinfo
  {journal} {J. Stat. Phys.}\ }\textbf {\bibinfo {volume} {180}},\ \bibinfo
  {pages} {474} (\bibinfo {year} {2020})}\BibitemShut {NoStop}%
\bibitem [{\citenamefont {Strating}(1999)}]{Strating:1999}%
  \BibitemOpen
  \bibfield  {author} {\bibinfo {author} {\bibfnamefont {P.}~\bibnamefont
  {Strating}},\ }\bibfield  {title} {\bibinfo {title} {Brownian dynamics
  simulation of a hard-sphere suspension},\ }\href
  {https://doi.org/10.1103/PhysRevE.59.2175} {\bibfield  {journal} {\bibinfo
  {journal} {Phys. Rev. E}\ }\textbf {\bibinfo {volume} {59}},\ \bibinfo
  {pages} {2175} (\bibinfo {year} {1999})}\BibitemShut {NoStop}%
\bibitem [{\citenamefont {Scala}(2012)}]{Scala:2012}%
  \BibitemOpen
  \bibfield  {author} {\bibinfo {author} {\bibfnamefont {A.}~\bibnamefont
  {Scala}},\ }\bibfield  {title} {\bibinfo {title} {Event-driven {L}angevin
  simulations of hard spheres},\ }\href
  {https://doi.org/10.1103/PhysRevE.86.026709} {\bibfield  {journal} {\bibinfo
  {journal} {Phys. Rev. E}\ }\textbf {\bibinfo {volume} {86}},\ \bibinfo
  {pages} {026709} (\bibinfo {year} {2012})}\BibitemShut {NoStop}%
\bibitem [{\citenamefont {Behringer}\ and\ \citenamefont
  {Eichhorn}(2012)}]{Behringer/Eichhorn:2012}%
  \BibitemOpen
  \bibfield  {author} {\bibinfo {author} {\bibfnamefont {H.}~\bibnamefont
  {Behringer}}\ and\ \bibinfo {author} {\bibfnamefont {R.}~\bibnamefont
  {Eichhorn}},\ }\bibfield  {title} {\bibinfo {title} {Brownian dynamics
  simulations with hard-body interactions: Spherical particles},\ }\href
  {https://doi.org/10.1063/1.4761827} {\bibfield  {journal} {\bibinfo
  {journal} {J. Chem. Phys.}\ }\textbf {\bibinfo {volume} {137}},\ \bibinfo
  {pages} {164108} (\bibinfo {year} {2012})}\BibitemShut {NoStop}%
\bibitem [{\citenamefont {{van Noije}}\ \emph {et~al.}(1998)\citenamefont {{van
  Noije}}, \citenamefont {Ernst},\ and\ \citenamefont
  {Brito}}]{Noije/etal:1998}%
  \BibitemOpen
  \bibfield  {author} {\bibinfo {author} {\bibfnamefont {T.~P.~C.}\
  \bibnamefont {{van Noije}}}, \bibinfo {author} {\bibfnamefont {M.~H.}\
  \bibnamefont {Ernst}},\ and\ \bibinfo {author} {\bibfnamefont
  {R.}~\bibnamefont {Brito}},\ }\bibfield  {title} {\bibinfo {title} {Ring
  kinetic theory for an idealized granular gas},\ }\href
  {https://doi.org/https://doi.org/10.1016/S0378-4371(97)00610-9} {\bibfield
  {journal} {\bibinfo  {journal} {Phys. A (Amsterdam, Neth.)}\ }\textbf
  {\bibinfo {volume} {251}},\ \bibinfo {pages} {266} (\bibinfo {year}
  {1998})}\BibitemShut {NoStop}%
\bibitem [{\citenamefont {Luding}(1998)}]{Luding:1998}%
  \BibitemOpen
  \bibfield  {author} {\bibinfo {author} {\bibfnamefont {S.}~\bibnamefont
  {Luding}},\ }\bibinfo {title} {Collisions {\&} contacts between two
  particles},\ in\ \href {https://doi.org/10.1007/978-94-017-2653-5_20} {\emph
  {\bibinfo {booktitle} {Physics of Dry Granular Media}}},\ \bibinfo {editor}
  {edited by\ \bibinfo {editor} {\bibfnamefont {H.~J.}\ \bibnamefont
  {Herrmann}}, \bibinfo {editor} {\bibfnamefont {J.-P.}\ \bibnamefont {Hovi}},\
  and\ \bibinfo {editor} {\bibfnamefont {S.}~\bibnamefont {Luding}}}\ (\bibinfo
   {publisher} {Springer Netherlands},\ \bibinfo {address} {Dordrecht},\
  \bibinfo {year} {1998})\ pp.\ \bibinfo {pages} {285--304}\BibitemShut
  {NoStop}%
\bibitem [{\citenamefont {Gonz{\'a}lez}\ \emph {et~al.}(2009)\citenamefont
  {Gonz{\'a}lez}, \citenamefont {Risso},\ and\ \citenamefont
  {Soto}}]{Gonzalez:2009}%
  \BibitemOpen
  \bibfield  {author} {\bibinfo {author} {\bibfnamefont {S.}~\bibnamefont
  {Gonz{\'a}lez}}, \bibinfo {author} {\bibfnamefont {D.}~\bibnamefont
  {Risso}},\ and\ \bibinfo {author} {\bibfnamefont {R.}~\bibnamefont {Soto}},\
  }\bibfield  {title} {\bibinfo {title} {Extended event driven molecular
  dynamics for simulating dense granular matter},\ }\href
  {https://doi.org/10.1140/epjst/e2010-01192-4} {\bibfield  {journal} {\bibinfo
   {journal} {Eur. Phys. J.: Spec. Top.}\ }\textbf {\bibinfo {volume} {179}},\
  \bibinfo {pages} {33} (\bibinfo {year} {2009})}\BibitemShut {NoStop}%
\bibitem [{\citenamefont {Helland}\ \emph {et~al.}(2007)\citenamefont
  {Helland}, \citenamefont {Bournot}, \citenamefont {Occelli},\ and\
  \citenamefont {Tadrist}}]{Helland/etal:2007}%
  \BibitemOpen
  \bibfield  {author} {\bibinfo {author} {\bibfnamefont {E.}~\bibnamefont
  {Helland}}, \bibinfo {author} {\bibfnamefont {H.}~\bibnamefont {Bournot}},
  \bibinfo {author} {\bibfnamefont {R.}~\bibnamefont {Occelli}},\ and\ \bibinfo
  {author} {\bibfnamefont {L.}~\bibnamefont {Tadrist}},\ }\bibfield  {title}
  {\bibinfo {title} {Drag reduction and cluster formation in a circulating
  fluidised bed},\ }\href
  {https://doi.org/https://doi.org/10.1016/j.ces.2006.08.012} {\bibfield
  {journal} {\bibinfo  {journal} {Chem. Eng. Sci.}\ }\textbf {\bibinfo {volume}
  {62}},\ \bibinfo {pages} {148} (\bibinfo {year} {2007})},\ \bibinfo {note}
  {fluidized Bed Applications}\BibitemShut {NoStop}%
\bibitem [{\citenamefont {McMillan}\ \emph {et~al.}(2013)\citenamefont
  {McMillan}, \citenamefont {Shaffer}, \citenamefont {Gopalan}, \citenamefont
  {Chew}, \citenamefont {Hrenya}, \citenamefont {Hays}, \citenamefont {Karri},\
  and\ \citenamefont {Cocco}}]{McMillan/etal:2013}%
  \BibitemOpen
  \bibfield  {author} {\bibinfo {author} {\bibfnamefont {J.}~\bibnamefont
  {McMillan}}, \bibinfo {author} {\bibfnamefont {F.}~\bibnamefont {Shaffer}},
  \bibinfo {author} {\bibfnamefont {B.}~\bibnamefont {Gopalan}}, \bibinfo
  {author} {\bibfnamefont {J.~W.}\ \bibnamefont {Chew}}, \bibinfo {author}
  {\bibfnamefont {C.}~\bibnamefont {Hrenya}}, \bibinfo {author} {\bibfnamefont
  {R.}~\bibnamefont {Hays}}, \bibinfo {author} {\bibfnamefont {S.~B.~R.}\
  \bibnamefont {Karri}},\ and\ \bibinfo {author} {\bibfnamefont
  {R.}~\bibnamefont {Cocco}},\ }\bibfield  {title} {\bibinfo {title} {Particle
  cluster dynamics during fluidization},\ }\href
  {https://doi.org/https://doi.org/10.1016/j.ces.2013.02.047} {\bibfield
  {journal} {\bibinfo  {journal} {Chem. Eng. Sci.}\ }\textbf {\bibinfo {volume}
  {100}},\ \bibinfo {pages} {39} (\bibinfo {year} {2013})},\ \bibinfo {note}
  {11th International Conference on Gas-Liquid and Gas-Liquid-Solid Reactor
  Engineering}\BibitemShut {NoStop}%
\bibitem [{\citenamefont {Cahyadi}\ \emph {et~al.}(2017)\citenamefont
  {Cahyadi}, \citenamefont {Anantharaman}, \citenamefont {Yang}, \citenamefont
  {Karri}, \citenamefont {Findlay}, \citenamefont {Cocco},\ and\ \citenamefont
  {Chew}}]{Cahyadi:2017}%
  \BibitemOpen
  \bibfield  {author} {\bibinfo {author} {\bibfnamefont {A.}~\bibnamefont
  {Cahyadi}}, \bibinfo {author} {\bibfnamefont {A.}~\bibnamefont
  {Anantharaman}}, \bibinfo {author} {\bibfnamefont {S.}~\bibnamefont {Yang}},
  \bibinfo {author} {\bibfnamefont {S.~R.}\ \bibnamefont {Karri}}, \bibinfo
  {author} {\bibfnamefont {J.~G.}\ \bibnamefont {Findlay}}, \bibinfo {author}
  {\bibfnamefont {R.~A.}\ \bibnamefont {Cocco}},\ and\ \bibinfo {author}
  {\bibfnamefont {J.~W.}\ \bibnamefont {Chew}},\ }\bibfield  {title} {\bibinfo
  {title} {Review of cluster characteristics in circulating fluidized bed (cfb)
  risers},\ }\href {https://doi.org/https://doi.org/10.1016/j.ces.2016.10.002}
  {\bibfield  {journal} {\bibinfo  {journal} {Chem. Eng. Sci.}\ }\textbf
  {\bibinfo {volume} {158}},\ \bibinfo {pages} {70} (\bibinfo {year}
  {2017})}\BibitemShut {NoStop}%
\bibitem [{\citenamefont {Lois}\ \emph {et~al.}(2006)\citenamefont {Lois},
  \citenamefont {Lema{\^{\i}}tre},\ and\ \citenamefont
  {Carlson}}]{Lois/etal:2006}%
  \BibitemOpen
  \bibfield  {author} {\bibinfo {author} {\bibfnamefont {G.}~\bibnamefont
  {Lois}}, \bibinfo {author} {\bibfnamefont {A.}~\bibnamefont
  {Lema{\^{\i}}tre}},\ and\ \bibinfo {author} {\bibfnamefont {J.~M.}\
  \bibnamefont {Carlson}},\ }\bibfield  {title} {\bibinfo {title} {Emergence of
  multi-contact interactions in contact dynamics simulations of granular shear
  flows},\ }\href {https://doi.org/10.1209/epl/i2005-10605-1} {\bibfield
  {journal} {\bibinfo  {journal} {Europhys. Lett. ({EPL})}\ }\textbf {\bibinfo
  {volume} {76}},\ \bibinfo {pages} {318} (\bibinfo {year} {2006})}\BibitemShut
  {NoStop}%
\bibitem [{\citenamefont {Watanabe}\ and\ \citenamefont
  {Tanaka}(2013)}]{Watanabe/Tanaka:2013}%
  \BibitemOpen
  \bibfield  {author} {\bibinfo {author} {\bibfnamefont {M.}~\bibnamefont
  {Watanabe}}\ and\ \bibinfo {author} {\bibfnamefont {D.}~\bibnamefont
  {Tanaka}},\ }\bibfield  {title} {\bibinfo {title} {Brownian dynamics
  simulation of the aggregation of submicron particles in static gas},\ }\href
  {https://doi.org/https://doi.org/10.1016/j.compchemeng.2013.03.028}
  {\bibfield  {journal} {\bibinfo  {journal} {Comput. Chem. Eng.}\ }\textbf
  {\bibinfo {volume} {54}},\ \bibinfo {pages} {151} (\bibinfo {year}
  {2013})}\BibitemShut {NoStop}%
\bibitem [{\citenamefont {Baxter}(1968)}]{Baxter:1968}%
  \BibitemOpen
  \bibfield  {author} {\bibinfo {author} {\bibfnamefont {R.~J.}\ \bibnamefont
  {Baxter}},\ }\bibfield  {title} {\bibinfo {title} {Percus--{Y}evick equation
  for hard spheres with surface adhesion},\ }\href
  {https://doi.org/10.1063/1.1670482} {\bibfield  {journal} {\bibinfo
  {journal} {J. Chem. Phys.}\ }\textbf {\bibinfo {volume} {49}},\ \bibinfo
  {pages} {2770} (\bibinfo {year} {1968})}\BibitemShut {NoStop}%
\bibitem [{\citenamefont {Percus}(1982)}]{Percus:1982}%
  \BibitemOpen
  \bibfield  {author} {\bibinfo {author} {\bibfnamefont {J.~K.}\ \bibnamefont
  {Percus}},\ }\bibfield  {title} {\bibinfo {title} {One-dimensional classical
  fluid with nearest-neighbor interaction in arbitrary external field},\ }\href
  {https://doi.org/10.1007/BF01011623} {\bibfield  {journal} {\bibinfo
  {journal} {J. Stat. Phys.}\ }\textbf {\bibinfo {volume} {28}},\ \bibinfo
  {pages} {67} (\bibinfo {year} {1982})}\BibitemShut {NoStop}%
\bibitem [{\citenamefont {Seaton}\ and\ \citenamefont
  {Glandt}(1987{\natexlab{a}})}]{Seaton/Glandt:1987_2}%
  \BibitemOpen
  \bibfield  {author} {\bibinfo {author} {\bibfnamefont {N.~A.}\ \bibnamefont
  {Seaton}}\ and\ \bibinfo {author} {\bibfnamefont {E.~D.}\ \bibnamefont
  {Glandt}},\ }\bibfield  {title} {\bibinfo {title} {{Monte Carlo} simulation
  of adhesive spheres},\ }\href {https://doi.org/10.1063/1.453724} {\bibfield
  {journal} {\bibinfo  {journal} {J. Chem. Phys.}\ }\textbf {\bibinfo {volume}
  {87}},\ \bibinfo {pages} {1785} (\bibinfo {year}
  {1987}{\natexlab{a}})}\BibitemShut {NoStop}%
\bibitem [{\citenamefont {Miller}\ and\ \citenamefont
  {Frenkel}(2004{\natexlab{a}})}]{Miller/Frenkel:2004_2}%
  \BibitemOpen
  \bibfield  {author} {\bibinfo {author} {\bibfnamefont {M.~A.}\ \bibnamefont
  {Miller}}\ and\ \bibinfo {author} {\bibfnamefont {D.}~\bibnamefont
  {Frenkel}},\ }\bibfield  {title} {\bibinfo {title} {Phase diagram of the
  adhesive hard sphere fluid},\ }\href {https://doi.org/10.1063/1.1758693}
  {\bibfield  {journal} {\bibinfo  {journal} {J. Chem. Phys.}\ }\textbf
  {\bibinfo {volume} {121}},\ \bibinfo {pages} {535} (\bibinfo {year}
  {2004}{\natexlab{a}})}\BibitemShut {NoStop}%
\bibitem [{\citenamefont {Miller}\ and\ \citenamefont
  {Frenkel}(2004{\natexlab{b}})}]{Miller/Frenkel:2004_1}%
  \BibitemOpen
  \bibfield  {author} {\bibinfo {author} {\bibfnamefont {M.~A.}\ \bibnamefont
  {Miller}}\ and\ \bibinfo {author} {\bibfnamefont {D.}~\bibnamefont
  {Frenkel}},\ }\bibfield  {title} {\bibinfo {title} {Simulating colloids with
  {Baxter's} adhesive hard sphere model},\ }\href
  {https://doi.org/10.1088/0953-8984/16/42/008} {\bibfield  {journal} {\bibinfo
   {journal} {J. Phys.: Condens. Matter}\ }\textbf {\bibinfo {volume} {16}},\
  \bibinfo {pages} {S4901} (\bibinfo {year} {2004}{\natexlab{b}})}\BibitemShut
  {NoStop}%
\bibitem [{\citenamefont {Foffi}\ \emph {et~al.}(2005)\citenamefont {Foffi},
  \citenamefont {Michele}, \citenamefont {Sciortino},\ and\ \citenamefont
  {Tartaglia}}]{Foffi/etal:2005}%
  \BibitemOpen
  \bibfield  {author} {\bibinfo {author} {\bibfnamefont {G.}~\bibnamefont
  {Foffi}}, \bibinfo {author} {\bibfnamefont {C.~D.}\ \bibnamefont {Michele}},
  \bibinfo {author} {\bibfnamefont {F.}~\bibnamefont {Sciortino}},\ and\
  \bibinfo {author} {\bibfnamefont {P.}~\bibnamefont {Tartaglia}},\ }\bibfield
  {title} {\bibinfo {title} {Scaling of dynamics with the range of interaction
  in short-range attractive colloids},\ }\href
  {https://doi.org/10.1103/PhysRevLett.94.078301} {\bibfield  {journal}
  {\bibinfo  {journal} {Phys. Rev. Lett.}\ }\textbf {\bibinfo {volume} {94}},\
  \bibinfo {pages} {078301} (\bibinfo {year} {2005})}\BibitemShut {NoStop}%
\bibitem [{\citenamefont {Zaccarelli}(2007)}]{Zaccarelli:2007}%
  \BibitemOpen
  \bibfield  {author} {\bibinfo {author} {\bibfnamefont {E.}~\bibnamefont
  {Zaccarelli}},\ }\bibfield  {title} {\bibinfo {title} {Colloidal gels:
  equilibrium and non-equilibrium routes},\ }\href
  {https://doi.org/10.1088/0953-8984/19/32/323101} {\bibfield  {journal}
  {\bibinfo  {journal} {J. Phys.: Condens. Matter}\ }\textbf {\bibinfo {volume}
  {19}},\ \bibinfo {pages} {323101} (\bibinfo {year} {2007})}\BibitemShut
  {NoStop}%
\bibitem [{\citenamefont {Wang}\ and\ \citenamefont
  {Swan}(2019)}]{Wang/Swan:2019}%
  \BibitemOpen
  \bibfield  {author} {\bibinfo {author} {\bibfnamefont {G.}~\bibnamefont
  {Wang}}\ and\ \bibinfo {author} {\bibfnamefont {J.~W.}\ \bibnamefont
  {Swan}},\ }\bibfield  {title} {\bibinfo {title} {Surface heterogeneity
  affects percolation and gelation of colloids: dynamic simulations with random
  patchy spheres},\ }\href {https://doi.org/10.1039/C9SM00607A} {\bibfield
  {journal} {\bibinfo  {journal} {Soft Matter}\ }\textbf {\bibinfo {volume}
  {15}},\ \bibinfo {pages} {5094} (\bibinfo {year} {2019})}\BibitemShut
  {NoStop}%
\bibitem [{\citenamefont {Whitaker}\ \emph {et~al.}(2019)\citenamefont
  {Whitaker}, \citenamefont {Varga}, \citenamefont {Hsiao}, \citenamefont
  {Solomon}, \citenamefont {Swan},\ and\ \citenamefont
  {Furst}}]{Whitaker/etal:2019}%
  \BibitemOpen
  \bibfield  {author} {\bibinfo {author} {\bibfnamefont {K.~A.}\ \bibnamefont
  {Whitaker}}, \bibinfo {author} {\bibfnamefont {Z.}~\bibnamefont {Varga}},
  \bibinfo {author} {\bibfnamefont {L.~C.}\ \bibnamefont {Hsiao}}, \bibinfo
  {author} {\bibfnamefont {M.~J.}\ \bibnamefont {Solomon}}, \bibinfo {author}
  {\bibfnamefont {J.~W.}\ \bibnamefont {Swan}},\ and\ \bibinfo {author}
  {\bibfnamefont {E.~M.}\ \bibnamefont {Furst}},\ }\bibfield  {title} {\bibinfo
  {title} {Colloidal gel elasticity arises from the packing of locally glassy
  clusters},\ }\href {https://doi.org/10.1038/s41467-019-10039-w} {\bibfield
  {journal} {\bibinfo  {journal} {Nat. Commun.}\ }\textbf {\bibinfo {volume}
  {10}},\ \bibinfo {pages} {2237} (\bibinfo {year} {2019})}\BibitemShut
  {NoStop}%
\bibitem [{\citenamefont {Stopper}\ \emph {et~al.}(2019)\citenamefont
  {Stopper}, \citenamefont {Hansen-Goos}, \citenamefont {Roth},\ and\
  \citenamefont {Evans}}]{Stopper/etal:2019}%
  \BibitemOpen
  \bibfield  {author} {\bibinfo {author} {\bibfnamefont {D.}~\bibnamefont
  {Stopper}}, \bibinfo {author} {\bibfnamefont {H.}~\bibnamefont
  {Hansen-Goos}}, \bibinfo {author} {\bibfnamefont {R.}~\bibnamefont {Roth}},\
  and\ \bibinfo {author} {\bibfnamefont {R.}~\bibnamefont {Evans}},\ }\bibfield
   {title} {\bibinfo {title} {On the decay of the pair correlation function and
  the line of vanishing excess isothermal compressibility in simple fluids},\
  }\href {https://doi.org/10.1063/1.5110044} {\bibfield  {journal} {\bibinfo
  {journal} {J. Chem. Phys.}\ }\textbf {\bibinfo {volume} {151}},\ \bibinfo
  {pages} {014501} (\bibinfo {year} {2019})}\BibitemShut {NoStop}%
\bibitem [{\citenamefont {Bou-Rabee}\ and\ \citenamefont
  {Holmes-Cerfon}(2020)}]{Bou-Rabee/etal:2020}%
  \BibitemOpen
  \bibfield  {author} {\bibinfo {author} {\bibfnamefont {N.}~\bibnamefont
  {Bou-Rabee}}\ and\ \bibinfo {author} {\bibfnamefont {M.~C.}\ \bibnamefont
  {Holmes-Cerfon}},\ }\bibfield  {title} {\bibinfo {title} {Sticky {Brownian}
  motion and its numerical solution},\ }\href
  {https://doi.org/10.1137/19M1268446} {\bibfield  {journal} {\bibinfo
  {journal} {SIAM Rev.}\ }\textbf {\bibinfo {volume} {62}},\ \bibinfo {pages}
  {164} (\bibinfo {year} {2020})}\BibitemShut {NoStop}%
\bibitem [{\citenamefont {Holmes-Cerfon}(2020)}]{Holmes-Cerfon:2020}%
  \BibitemOpen
  \bibfield  {author} {\bibinfo {author} {\bibfnamefont {M.}~\bibnamefont
  {Holmes-Cerfon}},\ }\bibfield  {title} {\bibinfo {title} {Simulating sticky
  particles: A {Monte Carlo} method to sample a stratification},\ }\href
  {https://doi.org/10.1063/5.0019550} {\bibfield  {journal} {\bibinfo
  {journal} {J. Chem. Phys.}\ }\textbf {\bibinfo {volume} {153}},\ \bibinfo
  {pages} {164112} (\bibinfo {year} {2020})}\BibitemShut {NoStop}%
\bibitem [{\citenamefont {Babu}\ \emph {et~al.}(2006)\citenamefont {Babu},
  \citenamefont {Gimel},\ and\ \citenamefont {Nicolai}}]{Babu/etal:2006}%
  \BibitemOpen
  \bibfield  {author} {\bibinfo {author} {\bibfnamefont {S.}~\bibnamefont
  {Babu}}, \bibinfo {author} {\bibfnamefont {J.~C.}\ \bibnamefont {Gimel}},\
  and\ \bibinfo {author} {\bibfnamefont {T.}~\bibnamefont {Nicolai}},\
  }\bibfield  {title} {\bibinfo {title} {Phase separation and percolation of
  reversibly aggregating spheres with a square-well attraction potential},\
  }\href {https://doi.org/10.1063/1.2378832} {\bibfield  {journal} {\bibinfo
  {journal} {J. Chem. Phys.}\ }\textbf {\bibinfo {volume} {125}},\ \bibinfo
  {pages} {184512} (\bibinfo {year} {2006})}\BibitemShut {NoStop}%
\bibitem [{\citenamefont {Kloeden}\ and\ \citenamefont
  {Platen}(1992)}]{Kloeden/Platen:1992}%
  \BibitemOpen
  \bibfield  {author} {\bibinfo {author} {\bibfnamefont {P.~E.}\ \bibnamefont
  {Kloeden}}\ and\ \bibinfo {author} {\bibfnamefont {E.}~\bibnamefont
  {Platen}},\ }\href
  {https://doi.org/https://doi.org/10.1007/978-3-662-12616-5} {\emph {\bibinfo
  {title} {Numerical Solution of Stochastic Differential Equations}}}\
  (\bibinfo  {publisher} {Springer Berlin, Heidelberg},\ \bibinfo {year}
  {1992})\BibitemShut {NoStop}%
\bibitem [{\citenamefont {Saito}\ and\ \citenamefont
  {Mitsui}(1993)}]{Saito/Mitsui:1993}%
  \BibitemOpen
  \bibfield  {author} {\bibinfo {author} {\bibfnamefont {Y.}~\bibnamefont
  {Saito}}\ and\ \bibinfo {author} {\bibfnamefont {T.}~\bibnamefont {Mitsui}},\
  }\bibfield  {title} {\bibinfo {title} {Simulation of stochastic differential
  equations},\ }\href {https://doi.org/10.1007/BF00773344} {\bibfield
  {journal} {\bibinfo  {journal} {Ann. Inst. Statist. Math.}\ }\textbf
  {\bibinfo {volume} {45}},\ \bibinfo {pages} {419} (\bibinfo {year}
  {1993})}\BibitemShut {NoStop}%
\bibitem [{\citenamefont {Lips}\ \emph {et~al.}(2018)\citenamefont {Lips},
  \citenamefont {Ryabov},\ and\ \citenamefont {Maass}}]{Lips/etal:2018}%
  \BibitemOpen
  \bibfield  {author} {\bibinfo {author} {\bibfnamefont {D.}~\bibnamefont
  {Lips}}, \bibinfo {author} {\bibfnamefont {A.}~\bibnamefont {Ryabov}},\ and\
  \bibinfo {author} {\bibfnamefont {P.}~\bibnamefont {Maass}},\ }\bibfield
  {title} {\bibinfo {title} {Brownian asymmetric simple exclusion process},\
  }\href {https://doi.org/10.1103/PhysRevLett.121.160601} {\bibfield  {journal}
  {\bibinfo  {journal} {Phys. Rev. Lett.}\ }\textbf {\bibinfo {volume} {121}},\
  \bibinfo {pages} {160601} (\bibinfo {year} {2018})}\BibitemShut {NoStop}%
\bibitem [{\citenamefont {Antonov}\ \emph {et~al.}(2022)\citenamefont
  {Antonov}, \citenamefont {Ryabov},\ and\ \citenamefont
  {Maass}}]{Antonov/etal:2022a}%
  \BibitemOpen
  \bibfield  {author} {\bibinfo {author} {\bibfnamefont {A.~P.}\ \bibnamefont
  {Antonov}}, \bibinfo {author} {\bibfnamefont {A.}~\bibnamefont {Ryabov}},\
  and\ \bibinfo {author} {\bibfnamefont {P.}~\bibnamefont {Maass}},\ }\bibfield
   {title} {\bibinfo {title} {Solitons in overdamped {B}rownian dynamics},\
  }\href {https://doi.org/10.1103/PhysRevLett.129.080601} {\bibfield  {journal}
  {\bibinfo  {journal} {Phys. Rev. Lett.}\ }\textbf {\bibinfo {volume} {129}},\
  \bibinfo {pages} {080601} (\bibinfo {year} {2022})}\BibitemShut {NoStop}%
\bibitem [{Ant()}]{Antonov/Schweers:2022}%
  \BibitemOpen
  \href@noop {} {}\bibinfo {note} {\Cpp{} implementation of the {B}rownian
  cluster dynamics ({BCD}) algorithm available from GitHub repository,
  \href{https://github.com/soeren802/Brownian-cluster-dynamics}{https://github.com/soeren802/Brownian-cluster-dynamics}}\BibitemShut
  {NoStop}%
\bibitem [{\citenamefont {Sanderson}\ and\ \citenamefont
  {Curtin}(2016)}]{Sanderson/Curtin:2016}%
  \BibitemOpen
  \bibfield  {author} {\bibinfo {author} {\bibfnamefont {C.}~\bibnamefont
  {Sanderson}}\ and\ \bibinfo {author} {\bibfnamefont {R.}~\bibnamefont
  {Curtin}},\ }\bibfield  {title} {\bibinfo {title} {Armadillo: a
  template-based {C}++ library for linear algebra},\ }\href
  {https://doi.org/10.21105/joss.00026} {\bibfield  {journal} {\bibinfo
  {journal} {J. Open Source Softw.}\ }\textbf {\bibinfo {volume} {1}},\
  \bibinfo {pages} {26} (\bibinfo {year} {2016})}\BibitemShut {NoStop}%
\bibitem [{\citenamefont {Sanderson}\ and\ \citenamefont
  {Curtin}(2018)}]{Sanderson/Curtin:2018}%
  \BibitemOpen
  \bibfield  {author} {\bibinfo {author} {\bibfnamefont {C.}~\bibnamefont
  {Sanderson}}\ and\ \bibinfo {author} {\bibfnamefont {R.}~\bibnamefont
  {Curtin}},\ }\bibfield  {title} {\bibinfo {title} {A user-friendly hybrid
  sparse matrix class in {C}++},\ }in\ \href
  {https://doi.org/10.1007/978-3-319-96418-8_50} {\emph {\bibinfo {booktitle}
  {Mathematical Software -- ICMS 2018}}},\ \bibinfo {editor} {edited by\
  \bibinfo {editor} {\bibfnamefont {J.~H.}\ \bibnamefont {Davenport}}, \bibinfo
  {editor} {\bibfnamefont {M.}~\bibnamefont {Kauers}}, \bibinfo {editor}
  {\bibfnamefont {G.}~\bibnamefont {Labahn}},\ and\ \bibinfo {editor}
  {\bibfnamefont {J.}~\bibnamefont {Urban}}}\ (\bibinfo  {publisher} {Springer
  International Publishing},\ \bibinfo {address} {Cham},\ \bibinfo {year}
  {2018})\ pp.\ \bibinfo {pages} {422--430}\BibitemShut {NoStop}%
\bibitem [{\citenamefont {Percus}(1976)}]{Percus:1976}%
  \BibitemOpen
  \bibfield  {author} {\bibinfo {author} {\bibfnamefont {J.~K.}\ \bibnamefont
  {Percus}},\ }\bibfield  {title} {\bibinfo {title} {Equilibrium state of a
  classical fluid of hard rods in an external field},\ }\href
  {https://doi.org/10.1007/BF01020803} {\bibfield  {journal} {\bibinfo
  {journal} {J. Stat. Phys.}\ }\textbf {\bibinfo {volume} {15}},\ \bibinfo
  {pages} {505} (\bibinfo {year} {1976})}\BibitemShut {NoStop}%
\bibitem [{\citenamefont {Marconi}\ and\ \citenamefont
  {Tarazona}(1999)}]{Marconi/Tarazona:1999}%
  \BibitemOpen
  \bibfield  {author} {\bibinfo {author} {\bibfnamefont {U.~M.~B.}\
  \bibnamefont {Marconi}}\ and\ \bibinfo {author} {\bibfnamefont
  {P.}~\bibnamefont {Tarazona}},\ }\bibfield  {title} {\bibinfo {title}
  {Dynamic density functional theory of fluids},\ }\href
  {https://doi.org/10.1063/1.478705} {\bibfield  {journal} {\bibinfo  {journal}
  {J. Chem. Phys.}\ }\textbf {\bibinfo {volume} {110}},\ \bibinfo {pages}
  {8032} (\bibinfo {year} {1999})}\BibitemShut {NoStop}%
\bibitem [{\citenamefont {te~Vrugt}\ \emph {et~al.}(2020)\citenamefont
  {te~Vrugt}, \citenamefont {L{\"o}wen},\ and\ \citenamefont
  {Wittkowski}}]{teVrugt/etal:2020}%
  \BibitemOpen
  \bibfield  {author} {\bibinfo {author} {\bibfnamefont {M.}~\bibnamefont
  {te~Vrugt}}, \bibinfo {author} {\bibfnamefont {H.}~\bibnamefont
  {L{\"o}wen}},\ and\ \bibinfo {author} {\bibfnamefont {R.}~\bibnamefont
  {Wittkowski}},\ }\bibfield  {title} {\bibinfo {title} {Classical dynamical
  density functional theory: from fundamentals to applications},\ }\href
  {https://doi.org/10.1080/00018732.2020.1854965} {\bibfield  {journal}
  {\bibinfo  {journal} {Adv. Phys.}\ }\textbf {\bibinfo {volume} {69}},\
  \bibinfo {pages} {121} (\bibinfo {year} {2020})}\BibitemShut {NoStop}%
\bibitem [{\citenamefont {Marconi}\ and\ \citenamefont
  {Tarazona}(2000)}]{Marconi/Tarazona:2000}%
  \BibitemOpen
  \bibfield  {author} {\bibinfo {author} {\bibfnamefont {U.~M.~B.}\
  \bibnamefont {Marconi}}\ and\ \bibinfo {author} {\bibfnamefont
  {P.}~\bibnamefont {Tarazona}},\ }\bibfield  {title} {\bibinfo {title}
  {Dynamic density functional theory of fluids},\ }\href
  {https://doi.org/10.1088/0953-8984/12/8a/356} {\bibfield  {journal} {\bibinfo
   {journal} {J. Phys.: Condens. Mat.}\ }\textbf {\bibinfo {volume} {12}},\
  \bibinfo {pages} {A413} (\bibinfo {year} {2000})}\BibitemShut {NoStop}%
\bibitem [{\citenamefont {Lewis}(2000)}]{Lewis:2000}%
  \BibitemOpen
  \bibfield  {author} {\bibinfo {author} {\bibfnamefont {J.~A.}\ \bibnamefont
  {Lewis}},\ }\bibfield  {title} {\bibinfo {title} {Colloidal processing of
  ceramics},\ }\href
  {https://doi.org/https://doi.org/10.1111/j.1151-2916.2000.tb01560.x}
  {\bibfield  {journal} {\bibinfo  {journal} {J. Am. Ceram. Soc.}\ }\textbf
  {\bibinfo {volume} {83}},\ \bibinfo {pages} {2341} (\bibinfo {year}
  {2000})}\BibitemShut {NoStop}%
\bibitem [{\citenamefont {Watts}(1969)}]{Watts:1969}%
  \BibitemOpen
  \bibfield  {author} {\bibinfo {author} {\bibfnamefont {R.~O.}\ \bibnamefont
  {Watts}},\ }\bibfield  {title} {\bibinfo {title} {Hypernetted‐chain
  approximation applied to a modified {Lennard‐Jones} fluid},\ }\href
  {https://doi.org/10.1063/1.1671198} {\bibfield  {journal} {\bibinfo
  {journal} {J. Chem. Phys.}\ }\textbf {\bibinfo {volume} {50}},\ \bibinfo
  {pages} {1358} (\bibinfo {year} {1969})}\BibitemShut {NoStop}%
\bibitem [{\citenamefont {Gonz{\'a}lez-Calder{\'o}n}\ \emph
  {et~al.}(2019)\citenamefont {Gonz{\'a}lez-Calder{\'o}n}, \citenamefont
  {Perera-Burgos},\ and\ \citenamefont {Luis}}]{Gonzalez-Calderon:2019}%
  \BibitemOpen
  \bibfield  {author} {\bibinfo {author} {\bibfnamefont {A.}~\bibnamefont
  {Gonz{\'a}lez-Calder{\'o}n}}, \bibinfo {author} {\bibfnamefont {J.~A.}\
  \bibnamefont {Perera-Burgos}},\ and\ \bibinfo {author} {\bibfnamefont
  {D.~P.}\ \bibnamefont {Luis}},\ }\bibfield  {title} {\bibinfo {title}
  {Critical temperatures of real fluids from the extended law of corresponding
  states},\ }\href {https://doi.org/10.1063/1.5123613} {\bibfield  {journal}
  {\bibinfo  {journal} {AIP Adv.}\ }\textbf {\bibinfo {volume} {9}},\ \bibinfo
  {pages} {115217} (\bibinfo {year} {2019})}\BibitemShut {NoStop}%
\bibitem [{\citenamefont {Trombach}\ \emph {et~al.}(2018)\citenamefont
  {Trombach}, \citenamefont {Hoy}, \citenamefont {Wales},\ and\ \citenamefont
  {Schwerdtfeger}}]{Trombach/etal:2018}%
  \BibitemOpen
  \bibfield  {author} {\bibinfo {author} {\bibfnamefont {L.}~\bibnamefont
  {Trombach}}, \bibinfo {author} {\bibfnamefont {R.~S.}\ \bibnamefont {Hoy}},
  \bibinfo {author} {\bibfnamefont {D.~J.}\ \bibnamefont {Wales}},\ and\
  \bibinfo {author} {\bibfnamefont {P.}~\bibnamefont {Schwerdtfeger}},\
  }\bibfield  {title} {\bibinfo {title} {From sticky-hard-sphere to
  {Lennard-Jones}-type clusters},\ }\href
  {https://doi.org/10.1103/PhysRevE.97.043309} {\bibfield  {journal} {\bibinfo
  {journal} {Phys. Rev. E}\ }\textbf {\bibinfo {volume} {97}},\ \bibinfo
  {pages} {043309} (\bibinfo {year} {2018})}\BibitemShut {NoStop}%
\bibitem [{\citenamefont {Chiew}\ and\ \citenamefont
  {Glandt}(1983)}]{Chiew/Glandt:1983}%
  \BibitemOpen
  \bibfield  {author} {\bibinfo {author} {\bibfnamefont {Y.~C.}\ \bibnamefont
  {Chiew}}\ and\ \bibinfo {author} {\bibfnamefont {E.~D.}\ \bibnamefont
  {Glandt}},\ }\bibfield  {title} {\bibinfo {title} {Percolation behaviour of
  permeable and of adhesive spheres},\ }\href
  {https://doi.org/10.1088/0305-4470/16/11/026} {\bibfield  {journal} {\bibinfo
   {journal} {J. Phys. A: Math. Gen.}\ }\textbf {\bibinfo {volume} {16}},\
  \bibinfo {pages} {2599} (\bibinfo {year} {1983})}\BibitemShut {NoStop}%
\bibitem [{\citenamefont {Seaton}\ and\ \citenamefont
  {Glandt}(1987{\natexlab{b}})}]{Seaton/Glandt:1987}%
  \BibitemOpen
  \bibfield  {author} {\bibinfo {author} {\bibfnamefont {N.~A.}\ \bibnamefont
  {Seaton}}\ and\ \bibinfo {author} {\bibfnamefont {E.~D.}\ \bibnamefont
  {Glandt}},\ }\bibfield  {title} {\bibinfo {title} {Aggregation and
  percolation in a system of adhesive spheres},\ }\href
  {https://doi.org/10.1063/1.452707} {\bibfield  {journal} {\bibinfo  {journal}
  {J. Chem. Phys.}\ }\textbf {\bibinfo {volume} {86}},\ \bibinfo {pages} {4668}
  (\bibinfo {year} {1987}{\natexlab{b}})}\BibitemShut {NoStop}%
\bibitem [{\citenamefont {Torquato}\ \emph {et~al.}(1988)\citenamefont
  {Torquato}, \citenamefont {Beasley},\ and\ \citenamefont
  {Chiew}}]{Torquato/etal:1988}%
  \BibitemOpen
  \bibfield  {author} {\bibinfo {author} {\bibfnamefont {S.}~\bibnamefont
  {Torquato}}, \bibinfo {author} {\bibfnamefont {J.~D.}\ \bibnamefont
  {Beasley}},\ and\ \bibinfo {author} {\bibfnamefont {Y.~C.}\ \bibnamefont
  {Chiew}},\ }\bibfield  {title} {\bibinfo {title} {Two‐point cluster
  function for continuum percolation},\ }\href
  {https://doi.org/10.1063/1.454440} {\bibfield  {journal} {\bibinfo  {journal}
  {J. Chem. Phys.}\ }\textbf {\bibinfo {volume} {88}},\ \bibinfo {pages} {6540}
  (\bibinfo {year} {1988})}\BibitemShut {NoStop}%
\bibitem [{\citenamefont {Kim}\ \emph {et~al.}(2014)\citenamefont {Kim},
  \citenamefont {Merger}, \citenamefont {Wilhelm},\ and\ \citenamefont
  {Helgeson}}]{Kim/etal:2014}%
  \BibitemOpen
  \bibfield  {author} {\bibinfo {author} {\bibfnamefont {J.}~\bibnamefont
  {Kim}}, \bibinfo {author} {\bibfnamefont {D.}~\bibnamefont {Merger}},
  \bibinfo {author} {\bibfnamefont {M.}~\bibnamefont {Wilhelm}},\ and\ \bibinfo
  {author} {\bibfnamefont {M.~E.}\ \bibnamefont {Helgeson}},\ }\bibfield
  {title} {\bibinfo {title} {Microstructure and nonlinear signatures of
  yielding in a heterogeneous colloidal gel under large amplitude oscillatory
  shear},\ }\href {https://doi.org/10.1122/1.4882019} {\bibfield  {journal}
  {\bibinfo  {journal} {J. Rheol. (Melville, NY, U. S.)}\ }\textbf {\bibinfo
  {volume} {58}},\ \bibinfo {pages} {1359} (\bibinfo {year}
  {2014})}\BibitemShut {NoStop}%
\bibitem [{\citenamefont {Piazza}(2014)}]{Piazza:2014}%
  \BibitemOpen
  \bibfield  {author} {\bibinfo {author} {\bibfnamefont {R.}~\bibnamefont
  {Piazza}},\ }\bibfield  {title} {\bibinfo {title} {Settled and unsettled
  issues in particle settling},\ }\href
  {https://doi.org/10.1088/0034-4885/77/5/056602} {\bibfield  {journal}
  {\bibinfo  {journal} {Rep. Prog. Phys.}\ }\textbf {\bibinfo {volume} {77}},\
  \bibinfo {pages} {056602} (\bibinfo {year} {2014})}\BibitemShut {NoStop}%
\bibitem [{\citenamefont {Balazs}\ \emph {et~al.}(2020)\citenamefont {Balazs},
  \citenamefont {Dunbar}, \citenamefont {Smilgies},\ and\ \citenamefont
  {Hanrath}}]{Balazs/etal:2020}%
  \BibitemOpen
  \bibfield  {author} {\bibinfo {author} {\bibfnamefont {D.~M.}\ \bibnamefont
  {Balazs}}, \bibinfo {author} {\bibfnamefont {T.~A.}\ \bibnamefont {Dunbar}},
  \bibinfo {author} {\bibfnamefont {D.-M.}\ \bibnamefont {Smilgies}},\ and\
  \bibinfo {author} {\bibfnamefont {T.}~\bibnamefont {Hanrath}},\ }\bibfield
  {title} {\bibinfo {title} {Coupled dynamics of colloidal nanoparticle
  spreading and self-assembly at a fluid--fluid interface},\ }\href
  {https://doi.org/10.1021/acs.langmuir.0c00524} {\bibfield  {journal}
  {\bibinfo  {journal} {Langmuir}\ }\textbf {\bibinfo {volume} {36}},\ \bibinfo
  {pages} {6106} (\bibinfo {year} {2020})},\ \bibinfo {note} {pMID:
  32390432}\BibitemShut {NoStop}%
\bibitem [{\citenamefont {Talbot}\ \emph {et~al.}(2000)\citenamefont {Talbot},
  \citenamefont {Tarjus}, \citenamefont {{Van Tassel}},\ and\ \citenamefont
  {Viot}}]{Talbot/etal:2000}%
  \BibitemOpen
  \bibfield  {author} {\bibinfo {author} {\bibfnamefont {J.}~\bibnamefont
  {Talbot}}, \bibinfo {author} {\bibfnamefont {G.}~\bibnamefont {Tarjus}},
  \bibinfo {author} {\bibfnamefont {P.}~\bibnamefont {{Van Tassel}}},\ and\
  \bibinfo {author} {\bibfnamefont {P.}~\bibnamefont {Viot}},\ }\bibfield
  {title} {\bibinfo {title} {From car parking to protein adsorption: An
  overview of sequential adsorption processes},\ }\href
  {https://doi.org/https://doi.org/10.1016/S0927-7757(99)00409-4} {\bibfield
  {journal} {\bibinfo  {journal} {Colloids Surf. A}\ }\textbf {\bibinfo
  {volume} {165}},\ \bibinfo {pages} {287} (\bibinfo {year}
  {2000})}\BibitemShut {NoStop}%
\bibitem [{\citenamefont {Richard}\ \emph {et~al.}(2018)\citenamefont
  {Richard}, \citenamefont {Hallett}, \citenamefont {Speck},\ and\
  \citenamefont {Royall}}]{Richard/etal:2018}%
  \BibitemOpen
  \bibfield  {author} {\bibinfo {author} {\bibfnamefont {D.}~\bibnamefont
  {Richard}}, \bibinfo {author} {\bibfnamefont {J.}~\bibnamefont {Hallett}},
  \bibinfo {author} {\bibfnamefont {T.}~\bibnamefont {Speck}},\ and\ \bibinfo
  {author} {\bibfnamefont {C.~P.}\ \bibnamefont {Royall}},\ }\bibfield  {title}
  {\bibinfo {title} {Coupling between criticality and gelation in ``sticky''
  spheres: a structural analysis},\ }\href {https://doi.org/10.1039/C8SM00389K}
  {\bibfield  {journal} {\bibinfo  {journal} {Soft Matter}\ }\textbf {\bibinfo
  {volume} {14}},\ \bibinfo {pages} {5554} (\bibinfo {year}
  {2018})}\BibitemShut {NoStop}%
\bibitem [{\citenamefont {Assenza}\ and\ \citenamefont
  {Mezzenga}(2019)}]{Assenza/Mezzenga:2019}%
  \BibitemOpen
  \bibfield  {author} {\bibinfo {author} {\bibfnamefont {S.}~\bibnamefont
  {Assenza}}\ and\ \bibinfo {author} {\bibfnamefont {R.}~\bibnamefont
  {Mezzenga}},\ }\bibfield  {title} {\bibinfo {title} {Soft condensed matter
  physics of foods and macronutrients},\ }\href
  {https://doi.org/10.1038/s42254-019-0077-8} {\bibfield  {journal} {\bibinfo
  {journal} {Nat. Rev. Phys.}\ }\textbf {\bibinfo {volume} {1}},\ \bibinfo
  {pages} {551} (\bibinfo {year} {2019})}\BibitemShut {NoStop}%
\bibitem [{\citenamefont {Smith}\ \emph {et~al.}(2020)\citenamefont {Smith},
  \citenamefont {Brok}, \citenamefont {Christiansen},\ and\ \citenamefont
  {Ahrn{\'e}}}]{Smith/etal:2020}%
  \BibitemOpen
  \bibfield  {author} {\bibinfo {author} {\bibfnamefont {G.~N.}\ \bibnamefont
  {Smith}}, \bibinfo {author} {\bibfnamefont {E.}~\bibnamefont {Brok}},
  \bibinfo {author} {\bibfnamefont {M.~V.}\ \bibnamefont {Christiansen}},\ and\
  \bibinfo {author} {\bibfnamefont {L.}~\bibnamefont {Ahrn{\'e}}},\ }\bibfield
  {title} {\bibinfo {title} {Casein micelles in milk as sticky spheres},\
  }\href {https://doi.org/10.1039/D0SM01327G} {\bibfield  {journal} {\bibinfo
  {journal} {Soft Matter}\ }\textbf {\bibinfo {volume} {16}},\ \bibinfo {pages}
  {9955} (\bibinfo {year} {2020})}\BibitemShut {NoStop}%
\bibitem [{\citenamefont {Rosenbaum}\ \emph {et~al.}(1996)\citenamefont
  {Rosenbaum}, \citenamefont {Zamora},\ and\ \citenamefont
  {Zukoski}}]{Rosenbaum/etal:1996}%
  \BibitemOpen
  \bibfield  {author} {\bibinfo {author} {\bibfnamefont {D.}~\bibnamefont
  {Rosenbaum}}, \bibinfo {author} {\bibfnamefont {P.~C.}\ \bibnamefont
  {Zamora}},\ and\ \bibinfo {author} {\bibfnamefont {C.~F.}\ \bibnamefont
  {Zukoski}},\ }\bibfield  {title} {\bibinfo {title} {Phase behavior of small
  attractive colloidal particles},\ }\href
  {https://doi.org/10.1103/PhysRevLett.76.150} {\bibfield  {journal} {\bibinfo
  {journal} {Phys. Rev. Lett.}\ }\textbf {\bibinfo {volume} {76}},\ \bibinfo
  {pages} {150} (\bibinfo {year} {1996})}\BibitemShut {NoStop}%
\bibitem [{\citenamefont {Genix}\ and\ \citenamefont
  {Oberdisse}(2018)}]{Genix/Oberdisse:2018}%
  \BibitemOpen
  \bibfield  {author} {\bibinfo {author} {\bibfnamefont {A.-C.}\ \bibnamefont
  {Genix}}\ and\ \bibinfo {author} {\bibfnamefont {J.}~\bibnamefont
  {Oberdisse}},\ }\bibfield  {title} {\bibinfo {title} {Nanoparticle
  self-assembly: From interactions in suspension to polymer nanocomposites},\
  }\href {https://doi.org/10.1039/C8SM00430G} {\bibfield  {journal} {\bibinfo
  {journal} {Soft Matter}\ }\textbf {\bibinfo {volume} {14}},\ \bibinfo {pages}
  {5161} (\bibinfo {year} {2018})}\BibitemShut {NoStop}%
\bibitem [{\citenamefont {Schwarz-Linek}\ \emph {et~al.}(2012)\citenamefont
  {Schwarz-Linek}, \citenamefont {Valeriani}, \citenamefont {Cacciuto},
  \citenamefont {Cates}, \citenamefont {Marenduzzo}, \citenamefont {Morozov},\
  and\ \citenamefont {Poon}}]{Schwarz-Linek/etal:2012}%
  \BibitemOpen
  \bibfield  {author} {\bibinfo {author} {\bibfnamefont {J.}~\bibnamefont
  {Schwarz-Linek}}, \bibinfo {author} {\bibfnamefont {C.}~\bibnamefont
  {Valeriani}}, \bibinfo {author} {\bibfnamefont {A.}~\bibnamefont {Cacciuto}},
  \bibinfo {author} {\bibfnamefont {M.~E.}\ \bibnamefont {Cates}}, \bibinfo
  {author} {\bibfnamefont {D.}~\bibnamefont {Marenduzzo}}, \bibinfo {author}
  {\bibfnamefont {A.~N.}\ \bibnamefont {Morozov}},\ and\ \bibinfo {author}
  {\bibfnamefont {W.~C.~K.}\ \bibnamefont {Poon}},\ }\bibfield  {title}
  {\bibinfo {title} {Phase separation and rotor self-assembly in active
  particle suspensions},\ }\href {https://doi.org/10.1073/pnas.1116334109}
  {\bibfield  {journal} {\bibinfo  {journal} {Proc. Natl. Acad. Sci. U. S. A.}\
  }\textbf {\bibinfo {volume} {109}},\ \bibinfo {pages} {4052} (\bibinfo {year}
  {2012})}\BibitemShut {NoStop}%
\bibitem [{\citenamefont {Arnoulx~de Pirey}\ \emph {et~al.}(2019)\citenamefont
  {Arnoulx~de Pirey}, \citenamefont {Lozano},\ and\ \citenamefont {van
  Wijland}}]{dePirey/etal:2019}%
  \BibitemOpen
  \bibfield  {author} {\bibinfo {author} {\bibfnamefont {T.}~\bibnamefont
  {Arnoulx~de Pirey}}, \bibinfo {author} {\bibfnamefont {G.}~\bibnamefont
  {Lozano}},\ and\ \bibinfo {author} {\bibfnamefont {F.}~\bibnamefont {van
  Wijland}},\ }\bibfield  {title} {\bibinfo {title} {Active hard spheres in
  infinitely many dimensions},\ }\href
  {https://doi.org/10.1103/PhysRevLett.123.260602} {\bibfield  {journal}
  {\bibinfo  {journal} {Phys. Rev. Lett.}\ }\textbf {\bibinfo {volume} {123}},\
  \bibinfo {pages} {260602} (\bibinfo {year} {2019})}\BibitemShut {NoStop}%
\bibitem [{\citenamefont {Bergenholtz}(2018)}]{Bergenholtz:2018}%
  \BibitemOpen
  \bibfield  {author} {\bibinfo {author} {\bibfnamefont {J.}~\bibnamefont
  {Bergenholtz}},\ }\bibfield  {title} {\bibinfo {title} {Detachment dynamics
  of colloidal spheres with adhesive interactions},\ }\href
  {https://doi.org/10.1103/PhysRevE.97.042610} {\bibfield  {journal} {\bibinfo
  {journal} {Phys. Rev. E}\ }\textbf {\bibinfo {volume} {97}},\ \bibinfo
  {pages} {042610} (\bibinfo {year} {2018})}\BibitemShut {NoStop}%
\bibitem [{\citenamefont {Wang}\ \emph {et~al.}(2019)\citenamefont {Wang},
  \citenamefont {Fiore},\ and\ \citenamefont {Swan}}]{Wang/etal:2019}%
  \BibitemOpen
  \bibfield  {author} {\bibinfo {author} {\bibfnamefont {G.}~\bibnamefont
  {Wang}}, \bibinfo {author} {\bibfnamefont {A.~M.}\ \bibnamefont {Fiore}},\
  and\ \bibinfo {author} {\bibfnamefont {J.~W.}\ \bibnamefont {Swan}},\
  }\bibfield  {title} {\bibinfo {title} {On the viscosity of adhesive hard
  sphere dispersions: Critical scaling and the role of rigid contacts},\ }\href
  {https://doi.org/10.1122/1.5063362} {\bibfield  {journal} {\bibinfo
  {journal} {J. Rheol. (Melville, NY, U. S.)}\ }\textbf {\bibinfo {volume}
  {63}},\ \bibinfo {pages} {229} (\bibinfo {year} {2019})}\BibitemShut
  {NoStop}%
\bibitem [{\citenamefont {von B{\"u}low}\ \emph {et~al.}(2019)\citenamefont
  {von B{\"u}low}, \citenamefont {Siggel}, \citenamefont {Linke},\ and\
  \citenamefont {Hummer}}]{vonBuelow/etal:2019}%
  \BibitemOpen
  \bibfield  {author} {\bibinfo {author} {\bibfnamefont {S.}~\bibnamefont {von
  B{\"u}low}}, \bibinfo {author} {\bibfnamefont {M.}~\bibnamefont {Siggel}},
  \bibinfo {author} {\bibfnamefont {M.}~\bibnamefont {Linke}},\ and\ \bibinfo
  {author} {\bibfnamefont {G.}~\bibnamefont {Hummer}},\ }\bibfield  {title}
  {\bibinfo {title} {Dynamic cluster formation determines viscosity and
  diffusion in dense protein solutions},\ }\href
  {https://doi.org/10.1073/pnas.1817564116} {\bibfield  {journal} {\bibinfo
  {journal} {Proc. Natl. Acad. Sci. U. S. A.}\ }\textbf {\bibinfo {volume}
  {116}},\ \bibinfo {pages} {9843} (\bibinfo {year} {2019})}\BibitemShut
  {NoStop}%
\bibitem [{\citenamefont {Bakhshandeh}\ \emph {et~al.}(2020)\citenamefont
  {Bakhshandeh}, \citenamefont {Frydel},\ and\ \citenamefont
  {Levin}}]{Bakhshandeh/etal:2020}%
  \BibitemOpen
  \bibfield  {author} {\bibinfo {author} {\bibfnamefont {A.}~\bibnamefont
  {Bakhshandeh}}, \bibinfo {author} {\bibfnamefont {D.}~\bibnamefont
  {Frydel}},\ and\ \bibinfo {author} {\bibfnamefont {Y.}~\bibnamefont
  {Levin}},\ }\bibfield  {title} {\bibinfo {title} {Charge regulation of
  colloidal particles in aqueous solutions},\ }\href
  {https://doi.org/10.1039/D0CP03633A} {\bibfield  {journal} {\bibinfo
  {journal} {Phys. Chem. Chem. Phys.}\ }\textbf {\bibinfo {volume} {22}},\
  \bibinfo {pages} {24712} (\bibinfo {year} {2020})}\BibitemShut {NoStop}%
\bibitem [{\citenamefont {Pedersen}(1997)}]{Pedersen:1997}%
  \BibitemOpen
  \bibfield  {author} {\bibinfo {author} {\bibfnamefont {J.~S.}\ \bibnamefont
  {Pedersen}},\ }\bibfield  {title} {\bibinfo {title} {Analysis of small-angle
  scattering data from colloids and polymer solutions: Modeling and
  least-squares fitting},\ }\href
  {https://doi.org/https://doi.org/10.1016/S0001-8686(97)00312-6} {\bibfield
  {journal} {\bibinfo  {journal} {Adv. Colloid Interface Sci.}\ }\textbf
  {\bibinfo {volume} {70}},\ \bibinfo {pages} {171} (\bibinfo {year}
  {1997})}\BibitemShut {NoStop}%
\bibitem [{\citenamefont {Svergun}\ and\ \citenamefont
  {Koch}(2003)}]{Svergun:2003}%
  \BibitemOpen
  \bibfield  {author} {\bibinfo {author} {\bibfnamefont {D.~I.}\ \bibnamefont
  {Svergun}}\ and\ \bibinfo {author} {\bibfnamefont {M.~H.~J.}\ \bibnamefont
  {Koch}},\ }\bibfield  {title} {\bibinfo {title} {Small-angle scattering
  studies of biological macromolecules in solution},\ }\href
  {https://doi.org/10.1088/0034-4885/66/10/r05} {\bibfield  {journal} {\bibinfo
   {journal} {Rep. Prog. Phys.}\ }\textbf {\bibinfo {volume} {66}},\ \bibinfo
  {pages} {1735} (\bibinfo {year} {2003})}\BibitemShut {NoStop}%
\bibitem [{\citenamefont {Li}\ \emph {et~al.}(2016)\citenamefont {Li},
  \citenamefont {Senesi},\ and\ \citenamefont {Lee}}]{Li/etal:2016}%
  \BibitemOpen
  \bibfield  {author} {\bibinfo {author} {\bibfnamefont {T.}~\bibnamefont
  {Li}}, \bibinfo {author} {\bibfnamefont {A.~J.}\ \bibnamefont {Senesi}},\
  and\ \bibinfo {author} {\bibfnamefont {B.}~\bibnamefont {Lee}},\ }\bibfield
  {title} {\bibinfo {title} {Small angle {X-ray} scattering for nanoparticle
  research},\ }\href {https://doi.org/10.1021/acs.chemrev.5b00690} {\bibfield
  {journal} {\bibinfo  {journal} {Chem. Rev.}\ }\textbf {\bibinfo {volume}
  {116}},\ \bibinfo {pages} {11128} (\bibinfo {year} {2016})},\ \bibinfo {note}
  {pMID: 27054962}\BibitemShut {NoStop}%
\bibitem [{\citenamefont {Motokawa}\ \emph {et~al.}(2019)\citenamefont
  {Motokawa}, \citenamefont {Kobayashi}, \citenamefont {Endo}, \citenamefont
  {Mu}, \citenamefont {Williams}, \citenamefont {Masters}, \citenamefont
  {Antonio}, \citenamefont {Heller},\ and\ \citenamefont
  {Nagao}}]{Motokawa/etal:2019}%
  \BibitemOpen
  \bibfield  {author} {\bibinfo {author} {\bibfnamefont {R.}~\bibnamefont
  {Motokawa}}, \bibinfo {author} {\bibfnamefont {T.}~\bibnamefont {Kobayashi}},
  \bibinfo {author} {\bibfnamefont {H.}~\bibnamefont {Endo}}, \bibinfo {author}
  {\bibfnamefont {J.}~\bibnamefont {Mu}}, \bibinfo {author} {\bibfnamefont
  {C.~D.}\ \bibnamefont {Williams}}, \bibinfo {author} {\bibfnamefont {A.~J.}\
  \bibnamefont {Masters}}, \bibinfo {author} {\bibfnamefont {M.~R.}\
  \bibnamefont {Antonio}}, \bibinfo {author} {\bibfnamefont {W.~T.}\
  \bibnamefont {Heller}},\ and\ \bibinfo {author} {\bibfnamefont
  {M.}~\bibnamefont {Nagao}},\ }\bibfield  {title} {\bibinfo {title} {A
  telescoping view of solute architectures in a complex fluid system},\ }\href
  {https://doi.org/10.1021/acscentsci.8b00669} {\bibfield  {journal} {\bibinfo
  {journal} {ACS Cent. Sci.}\ }\textbf {\bibinfo {volume} {5}},\ \bibinfo
  {pages} {85} (\bibinfo {year} {2019})},\ \bibinfo {note} {pMID:
  30693328}\BibitemShut {NoStop}%
\bibitem [{\citenamefont {Ko}\ \emph {et~al.}(2021)\citenamefont {Ko},
  \citenamefont {Henschel}, \citenamefont {Meledam}, \citenamefont {Schroer},
  \citenamefont {M{\"u}ller-Buschbaum}, \citenamefont {Laschewsky},\ and\
  \citenamefont {Papadakis}}]{Ko/etal:2021}%
  \BibitemOpen
  \bibfield  {author} {\bibinfo {author} {\bibfnamefont {C.-H.}\ \bibnamefont
  {Ko}}, \bibinfo {author} {\bibfnamefont {C.}~\bibnamefont {Henschel}},
  \bibinfo {author} {\bibfnamefont {G.~P.}\ \bibnamefont {Meledam}}, \bibinfo
  {author} {\bibfnamefont {M.~A.}\ \bibnamefont {Schroer}}, \bibinfo {author}
  {\bibfnamefont {P.}~\bibnamefont {M{\"u}ller-Buschbaum}}, \bibinfo {author}
  {\bibfnamefont {A.}~\bibnamefont {Laschewsky}},\ and\ \bibinfo {author}
  {\bibfnamefont {C.~M.}\ \bibnamefont {Papadakis}},\ }\bibfield  {title}
  {\bibinfo {title} {Self-assembled micelles from thermoresponsive poly(methyl
  methacrylate)-b-poly(n-isopropylacrylamide) diblock copolymers in aqueous
  solution},\ }\href {https://doi.org/10.1021/acs.macromol.0c02189} {\bibfield
  {journal} {\bibinfo  {journal} {Macromolecules}\ }\textbf {\bibinfo {volume}
  {54}},\ \bibinfo {pages} {384} (\bibinfo {year} {2021})}\BibitemShut
  {NoStop}%
\bibitem [{\citenamefont {Heil}\ and\ \citenamefont
  {Jayaraman}(2021)}]{Heil/Jayaraman:2021}%
  \BibitemOpen
  \bibfield  {author} {\bibinfo {author} {\bibfnamefont {C.~M.}\ \bibnamefont
  {Heil}}\ and\ \bibinfo {author} {\bibfnamefont {A.}~\bibnamefont
  {Jayaraman}},\ }\bibfield  {title} {\bibinfo {title} {Computational
  reverse-engineering analysis for scattering experiments of assembled binary
  mixture of nanoparticles},\ }\href
  {https://doi.org/10.1021/acsmaterialsau.1c00015} {\bibfield  {journal}
  {\bibinfo  {journal} {ACS Materials Au}\ }\textbf {\bibinfo {volume} {1}},\
  \bibinfo {pages} {140} (\bibinfo {year} {2021})}\BibitemShut {NoStop}%
\bibitem [{\citenamefont {Jeffries}\ \emph {et~al.}(2021)\citenamefont
  {Jeffries}, \citenamefont {Ilavsky}, \citenamefont {Martel}, \citenamefont
  {Hinrichs}, \citenamefont {Meyer}, \citenamefont {Pedersen}, \citenamefont
  {Sokolova},\ and\ \citenamefont {Svergun}}]{Jeffries/etal:2021}%
  \BibitemOpen
  \bibfield  {author} {\bibinfo {author} {\bibfnamefont {C.~M.}\ \bibnamefont
  {Jeffries}}, \bibinfo {author} {\bibfnamefont {J.}~\bibnamefont {Ilavsky}},
  \bibinfo {author} {\bibfnamefont {A.}~\bibnamefont {Martel}}, \bibinfo
  {author} {\bibfnamefont {S.}~\bibnamefont {Hinrichs}}, \bibinfo {author}
  {\bibfnamefont {A.}~\bibnamefont {Meyer}}, \bibinfo {author} {\bibfnamefont
  {J.~S.}\ \bibnamefont {Pedersen}}, \bibinfo {author} {\bibfnamefont {A.~V.}\
  \bibnamefont {Sokolova}},\ and\ \bibinfo {author} {\bibfnamefont {D.~I.}\
  \bibnamefont {Svergun}},\ }\bibfield  {title} {\bibinfo {title} {Small-angle
  {X-ray} and neutron scattering},\ }\href
  {https://doi.org/10.1038/s43586-021-00064-9} {\bibfield  {journal} {\bibinfo
  {journal} {Nat. Rev. Methods Primers}\ }\textbf {\bibinfo {volume} {1}},\
  \bibinfo {pages} {70} (\bibinfo {year} {2021})}\BibitemShut {NoStop}%
\bibitem [{\citenamefont {Heil}\ \emph {et~al.}(2022)\citenamefont {Heil},
  \citenamefont {Patil}, \citenamefont {Dhinojwala},\ and\ \citenamefont
  {Jayaraman}}]{Heil/etal:2022}%
  \BibitemOpen
  \bibfield  {author} {\bibinfo {author} {\bibfnamefont {C.~M.}\ \bibnamefont
  {Heil}}, \bibinfo {author} {\bibfnamefont {A.}~\bibnamefont {Patil}},
  \bibinfo {author} {\bibfnamefont {A.}~\bibnamefont {Dhinojwala}},\ and\
  \bibinfo {author} {\bibfnamefont {A.}~\bibnamefont {Jayaraman}},\ }\bibfield
  {title} {\bibinfo {title} {Computational reverse-engineering analysis for
  scattering experiments {(CREASE)} with machine learning enhancement to
  determine structure of nanoparticle mixtures and solutions},\ }\href
  {https://doi.org/10.1021/acscentsci.2c00382} {\bibfield  {journal} {\bibinfo
  {journal} {ACS Cent. Sci.}\ }\textbf {\bibinfo {volume} {8}},\ \bibinfo
  {pages} {996} (\bibinfo {year} {2022})}\BibitemShut {NoStop}%
\bibitem [{\citenamefont {Bergenholtz}\ and\ \citenamefont
  {Fuchs}(1999)}]{Bergenholtz/Fuchs:1999}%
  \BibitemOpen
  \bibfield  {author} {\bibinfo {author} {\bibfnamefont {J.}~\bibnamefont
  {Bergenholtz}}\ and\ \bibinfo {author} {\bibfnamefont {M.}~\bibnamefont
  {Fuchs}},\ }\bibfield  {title} {\bibinfo {title} {Nonergodicity transitions
  in colloidal suspensions with attractive interactions},\ }\href
  {https://doi.org/10.1103/PhysRevE.59.5706} {\bibfield  {journal} {\bibinfo
  {journal} {Phys. Rev. E}\ }\textbf {\bibinfo {volume} {59}},\ \bibinfo
  {pages} {5706} (\bibinfo {year} {1999})}\BibitemShut {NoStop}%
\bibitem [{\citenamefont {Dawson}\ \emph {et~al.}(2000)\citenamefont {Dawson},
  \citenamefont {Foffi}, \citenamefont {Fuchs}, \citenamefont {G\"otze},
  \citenamefont {Sciortino}, \citenamefont {Sperl}, \citenamefont {Tartaglia},
  \citenamefont {Voigtmann},\ and\ \citenamefont
  {Zaccarelli}}]{Dawson/etal:2000}%
  \BibitemOpen
  \bibfield  {author} {\bibinfo {author} {\bibfnamefont {K.}~\bibnamefont
  {Dawson}}, \bibinfo {author} {\bibfnamefont {G.}~\bibnamefont {Foffi}},
  \bibinfo {author} {\bibfnamefont {M.}~\bibnamefont {Fuchs}}, \bibinfo
  {author} {\bibfnamefont {W.}~\bibnamefont {G\"otze}}, \bibinfo {author}
  {\bibfnamefont {F.}~\bibnamefont {Sciortino}}, \bibinfo {author}
  {\bibfnamefont {M.}~\bibnamefont {Sperl}}, \bibinfo {author} {\bibfnamefont
  {P.}~\bibnamefont {Tartaglia}}, \bibinfo {author} {\bibfnamefont
  {T.}~\bibnamefont {Voigtmann}},\ and\ \bibinfo {author} {\bibfnamefont
  {E.}~\bibnamefont {Zaccarelli}},\ }\bibfield  {title} {\bibinfo {title}
  {Higher-order glass-transition singularities in colloidal systems with
  attractive interactions},\ }\href
  {https://doi.org/10.1103/PhysRevE.63.011401} {\bibfield  {journal} {\bibinfo
  {journal} {Phys. Rev. E}\ }\textbf {\bibinfo {volume} {63}},\ \bibinfo
  {pages} {011401} (\bibinfo {year} {2000})}\BibitemShut {NoStop}%
\bibitem [{\citenamefont {Parisi}\ and\ \citenamefont
  {Zamponi}(2010)}]{Parisi/Zamponi:2010}%
  \BibitemOpen
  \bibfield  {author} {\bibinfo {author} {\bibfnamefont {G.}~\bibnamefont
  {Parisi}}\ and\ \bibinfo {author} {\bibfnamefont {F.}~\bibnamefont
  {Zamponi}},\ }\bibfield  {title} {\bibinfo {title} {Mean-field theory of hard
  sphere glasses and jamming},\ }\href
  {https://doi.org/10.1103/RevModPhys.82.789} {\bibfield  {journal} {\bibinfo
  {journal} {Rev. Mod. Phys.}\ }\textbf {\bibinfo {volume} {82}},\ \bibinfo
  {pages} {789} (\bibinfo {year} {2010})}\BibitemShut {NoStop}%
\bibitem [{\citenamefont {Fullerton}\ and\ \citenamefont
  {Berthier}(2020)}]{Fullerton/Berthier:2020}%
  \BibitemOpen
  \bibfield  {author} {\bibinfo {author} {\bibfnamefont {C.~J.}\ \bibnamefont
  {Fullerton}}\ and\ \bibinfo {author} {\bibfnamefont {L.}~\bibnamefont
  {Berthier}},\ }\bibfield  {title} {\bibinfo {title} {Glassy behavior of
  sticky spheres: What lies beyond experimental timescales?},\ }\href
  {https://doi.org/10.1103/PhysRevLett.125.258004} {\bibfield  {journal}
  {\bibinfo  {journal} {Phys. Rev. Lett.}\ }\textbf {\bibinfo {volume} {125}},\
  \bibinfo {pages} {258004} (\bibinfo {year} {2020})}\BibitemShut {NoStop}%
\bibitem [{\citenamefont {Baxter}(1970)}]{Baxter:1970}%
  \BibitemOpen
  \bibfield  {author} {\bibinfo {author} {\bibfnamefont {R.~J.}\ \bibnamefont
  {Baxter}},\ }\bibfield  {title} {\bibinfo {title} {{Ornstein--Zernike}
  relation and {Percus--Yevick} approximation for fluid mixtures},\ }\href
  {https://doi.org/10.1063/1.1673684} {\bibfield  {journal} {\bibinfo
  {journal} {J. Chem. Phys.}\ }\textbf {\bibinfo {volume} {52}},\ \bibinfo
  {pages} {4559} (\bibinfo {year} {1970})}\BibitemShut {NoStop}%
\bibitem [{\citenamefont {Jamnik}(2008)}]{Jamnik:2008}%
  \BibitemOpen
  \bibfield  {author} {\bibinfo {author} {\bibfnamefont {A.}~\bibnamefont
  {Jamnik}},\ }\bibfield  {title} {\bibinfo {title} {Simulating asymmetric
  colloidal mixture with adhesive hard sphere model},\ }\href
  {https://doi.org/10.1063/1.2939120} {\bibfield  {journal} {\bibinfo
  {journal} {J. Chem. Phys.}\ }\textbf {\bibinfo {volume} {128}},\ \bibinfo
  {pages} {234504} (\bibinfo {year} {2008})}\BibitemShut {NoStop}%
\bibitem [{\citenamefont {Opdam}\ \emph {et~al.}(2021)\citenamefont {Opdam},
  \citenamefont {Schelling},\ and\ \citenamefont {Tuinier}}]{Opdam/etal:2021}%
  \BibitemOpen
  \bibfield  {author} {\bibinfo {author} {\bibfnamefont {J.}~\bibnamefont
  {Opdam}}, \bibinfo {author} {\bibfnamefont {M.~P.~M.}\ \bibnamefont
  {Schelling}},\ and\ \bibinfo {author} {\bibfnamefont {R.}~\bibnamefont
  {Tuinier}},\ }\bibfield  {title} {\bibinfo {title} {Phase behavior of binary
  hard-sphere mixtures: Free volume theory including reservoir hard-core
  interactions},\ }\href {https://doi.org/10.1063/5.0037963} {\bibfield
  {journal} {\bibinfo  {journal} {J. Chem. Phys.}\ }\textbf {\bibinfo {volume}
  {154}},\ \bibinfo {pages} {074902} (\bibinfo {year} {2021})}\BibitemShut
  {NoStop}%
\bibitem [{\citenamefont {Kobayashi}\ \emph {et~al.}(2021)\citenamefont
  {Kobayashi}, \citenamefont {Rohrbach}, \citenamefont {Scheichl},
  \citenamefont {Wilding},\ and\ \citenamefont {Jack}}]{Kobayashi/etal:2021}%
  \BibitemOpen
  \bibfield  {author} {\bibinfo {author} {\bibfnamefont {H.}~\bibnamefont
  {Kobayashi}}, \bibinfo {author} {\bibfnamefont {P.~B.}\ \bibnamefont
  {Rohrbach}}, \bibinfo {author} {\bibfnamefont {R.}~\bibnamefont {Scheichl}},
  \bibinfo {author} {\bibfnamefont {N.~B.}\ \bibnamefont {Wilding}},\ and\
  \bibinfo {author} {\bibfnamefont {R.~L.}\ \bibnamefont {Jack}},\ }\bibfield
  {title} {\bibinfo {title} {Critical point for demixing of binary hard
  spheres},\ }\href {https://doi.org/10.1103/PhysRevE.104.044603} {\bibfield
  {journal} {\bibinfo  {journal} {Phys. Rev. E}\ }\textbf {\bibinfo {volume}
  {104}},\ \bibinfo {pages} {044603} (\bibinfo {year} {2021})}\BibitemShut
  {NoStop}%
\bibitem [{\citenamefont {Zhang}\ \emph {et~al.}(2019)\citenamefont {Zhang},
  \citenamefont {Travitz},\ and\ \citenamefont {Larson}}]{Zhang/etal:2019}%
  \BibitemOpen
  \bibfield  {author} {\bibinfo {author} {\bibfnamefont {W.}~\bibnamefont
  {Zhang}}, \bibinfo {author} {\bibfnamefont {A.}~\bibnamefont {Travitz}},\
  and\ \bibinfo {author} {\bibfnamefont {R.~G.}\ \bibnamefont {Larson}},\
  }\bibfield  {title} {\bibinfo {title} {Modeling intercolloidal interactions
  induced by adsorption of mobile telechelic polymers onto particle surfaces},\
  }\href {https://doi.org/10.1021/acs.macromol.9b00775} {\bibfield  {journal}
  {\bibinfo  {journal} {Macromolecules}\ }\textbf {\bibinfo {volume} {52}},\
  \bibinfo {pages} {5357} (\bibinfo {year} {2019})}\BibitemShut {NoStop}%
\bibitem [{\citenamefont {Amani}\ \emph {et~al.}(2021)\citenamefont {Amani},
  \citenamefont {Karakashev}, \citenamefont {Grozev}, \citenamefont
  {Simeonova}, \citenamefont {Miller}, \citenamefont {Rudolph},\ and\
  \citenamefont {Firouzi}}]{Amani/etal:2021}%
  \BibitemOpen
  \bibfield  {author} {\bibinfo {author} {\bibfnamefont {P.}~\bibnamefont
  {Amani}}, \bibinfo {author} {\bibfnamefont {S.~I.}\ \bibnamefont
  {Karakashev}}, \bibinfo {author} {\bibfnamefont {N.~A.}\ \bibnamefont
  {Grozev}}, \bibinfo {author} {\bibfnamefont {S.~S.}\ \bibnamefont
  {Simeonova}}, \bibinfo {author} {\bibfnamefont {R.}~\bibnamefont {Miller}},
  \bibinfo {author} {\bibfnamefont {V.}~\bibnamefont {Rudolph}},\ and\ \bibinfo
  {author} {\bibfnamefont {M.}~\bibnamefont {Firouzi}},\ }\bibfield  {title}
  {\bibinfo {title} {Effect of selected monovalent salts on surfactant
  stabilized foams},\ }\href
  {https://doi.org/https://doi.org/10.1016/j.cis.2021.102490} {\bibfield
  {journal} {\bibinfo  {journal} {Adv. Colloid Interface Sci.}\ }\textbf
  {\bibinfo {volume} {295}},\ \bibinfo {pages} {102490} (\bibinfo {year}
  {2021})}\BibitemShut {NoStop}%
\bibitem [{\citenamefont {Bye}\ and\ \citenamefont
  {Curtis}(2019)}]{Bye/Curtis:2019}%
  \BibitemOpen
  \bibfield  {author} {\bibinfo {author} {\bibfnamefont {J.~W.}\ \bibnamefont
  {Bye}}\ and\ \bibinfo {author} {\bibfnamefont {R.~A.}\ \bibnamefont
  {Curtis}},\ }\bibfield  {title} {\bibinfo {title} {Controlling phase
  separation of lysozyme with polyvalent anions},\ }\href
  {https://doi.org/10.1021/acs.jpcb.8b10868} {\bibfield  {journal} {\bibinfo
  {journal} {J. Phys. Chem. B}\ }\textbf {\bibinfo {volume} {123}},\ \bibinfo
  {pages} {593} (\bibinfo {year} {2019})}\BibitemShut {NoStop}%
\bibitem [{Note1()}]{Note1}%
  \BibitemOpen
  \bibinfo {note} {The computational time for generating the stationary profile
  in the DDFT increases as $\Delta x^{-2}$. Our choice $\Delta x=10^{-3}$
  allowed us to obtain the equilibrium profile in a computational time of about
  one day with an Intel Core i5-7600 CPU 3.50GHz processor. For a spatial
  resolution $\Delta x=10^{-4}$ the computational time would be 100
  days.}\BibitemShut {Stop}%
\end{thebibliography}

%Control: production of eprint (1) enabled
%

\end{document}